\documentclass[aps,prd,notitlepage,showpacs,nofootinbib,superscriptaddress,12pt]{revtex4-1}

\usepackage[utf8]{inputenc} 

\usepackage{amsmath}
\usepackage{amssymb}
\usepackage{float}
\usepackage{comment}
\usepackage{slashed}
\usepackage[normalem]{ulem}
\usepackage{bbold}
\usepackage{cancel}
\usepackage{graphicx}
\usepackage{multirow}
\usepackage{rotating}
\usepackage{lipsum}
\usepackage{caption}
\usepackage{subcaption}
\usepackage{tabulary}

\newcolumntype{K}[1]{>{\centering\arraybackslash}m{#1}}
\def\gsim{\raise0.3ex\hbox{$\;>$\kern-0.75em\raise-1.1ex\hbox{$\sim\;$}}}
\def\lsim{\raise0.3ex\hbox{$\;<$\kern-0.75em\raise-1.1ex\hbox{$\sim\;$}}}

\usepackage[usenames,dvipsnames]{color}

\newcommand {\ignore}[1]{}

\usepackage[colorinlistoftodos]{todonotes}
\usepackage[colorlinks=true,linkcolor=darkred,citecolor=darkred,urlcolor=darkred, pdfborder={0 0 0}]{hyperref}
\usepackage[normalem]{ulem}

\usepackage{caption}
\usepackage{subcaption}
\definecolor{linkcolor}{rgb}{0,0,0.5}

\def\rd#1{\textcolor{red}{#1}}

\definecolor{darkgreen}{rgb}{0,0.5,0}
\definecolor{darkred}{rgb}{0.6,0,0}
\definecolor{brown}{rgb}{0.59, 0.29, 0.0}
\definecolor{mightnightblue}{RGB}{25,25,112}

\def\SM{$\text{SU}(3)_c \otimes \text{SU}(2)_L \otimes \text{U}(1)_Y$ }

\def\Y{\mathbf{Y}}

\newcommand{\AddrCFTP}{%
Departamento de F\'{\i}sica and CFTP, Instituto Superior T\'ecnico, Universidade de Lisboa, Av. Rovisco Pais 1, 1049-001 Lisboa, Portugal}

\newcommand{\AddrBhopal}{Department of Physics, Indian Institute of Science Education and Research - Bhopal, Bhopal Bypass Road, Bhauri, Bhopal 462066, India}

\newcommand{\AddrAHEP}{%
  AHEP Group, Institut de F\'{i}sica Corpuscular --
  CSIC/Universitat de Val\`{e}ncia, Parc Cient\'ific de Paterna.\\
 C/ Catedr\'atico Jos\'e Beltr\'an, 2 E-46980 Paterna (Valencia) - SPAIN}

\newcommand{\AddrIITBHU}{ Department of Physics, Indian Institute of Technology (BHU), Varanasi 221005, India}

\begin{document}

\title{ \color{BrickRed} 
Axion framework with color-mediated Dirac neutrino masses
}

\author{A. Batra}\email{aditya.batra@tecnico.ulisboa.pt}
\affiliation{\AddrCFTP}

\author{H.~B. C\^amara}\email{henrique.b.camara@tecnico.ulisboa.pt}
\affiliation{\AddrCFTP}

\author{F.~R. Joaquim}\email{filipe.joaquim@tecnico.ulisboa.pt}
\affiliation{\AddrCFTP}

\author{N.~Nath}\email{nnath.phy@iitbhu.ac.in} \affiliation{\AddrIITBHU}  \affiliation{\AddrAHEP}

\author{R. Srivastava}\email{rahul@iiserb.ac.in}
\affiliation{\AddrBhopal}

\author{J.~W.~F. Valle}\email{valle@ific.uv.es}
\affiliation{\AddrAHEP}

\begin{abstract}
\vspace{+0.5cm}
{\noindent
We propose a KSVZ-type axion framework in which vector-like quarks (VLQ) and colored scalars  generate Dirac neutrino masses radiatively. The global Peccei-Quinn symmetry (under which the exotic fermions are charged) addresses the strong CP problem and ensures the Dirac nature of neutrinos. The axion also accounts for the observed cosmological dark matter. We systematically explore all viable VLQ representations. Depending on the specific scenario, the framework predicts distinct axion-to-photon couplings, testable through haloscope and helioscope experiments, as well as potentially significant flavor-violating quark-axion interactions.
}
\end{abstract}

\maketitle

\section{Introduction}
\label{sec:intro}

Since the groundbreaking discovery of neutrino oscillation phenomena ~\cite{Kajita:2016cak,McDonald:2016ixn}, the pursuit of a deeper understanding of neutrino properties has paved a well-defined path toward exploring physics beyond the Standard Model (SM). 
Within the framework of the SM, neutrinos are massless, and lepton mixing does not occur.
Unlike other SM fermions, neutrinos have the unique potential to be their own antiparticles~\cite{Majorana:1937vz}. 
Indeed, as the only truly neutral fermions in nature, gauge theories suggest that if neutrinos are massive, they should be of the Majorana type~\cite{Schechter:1980gr}.
This intriguing possibility has inspired a wealth of theoretical models for Majorana neutrino mass generation. These models implement the seesaw mechanisms~\cite{Minkowski:1977sc,Gell-Mann:1979vob,Yanagida:1979as,Schechter:1980gr,Glashow:1979nm,Mohapatra:1979ia} and radiative mass generation schemes~\cite{Zee:1985id,Babu:1988ki,Tao:1996vb,Ma:2006km}, all of which serve as ultraviolet (UV) complete realizations of the effective Weinberg operator~\cite{Weinberg:1980wa}.
Experimentally, investigations into neutrinoless double beta decay~($0\nu\beta\beta$) offer a pathway to determine the fundamental nature of neutrinos—specifically, whether they are Dirac or Majorana particles. According to the black-box theorem, a positive observation of $0\nu\beta\beta$ would confirm that at least one neutrino is a Majorana particle~\cite{Schechter:1981bd}.
However, so far, $0\nu\beta\beta$ decay has not been observed~\cite{Jones:2021cga,Cirigliano:2022oqy,Dolinski:2019nrj}. This could very well indicate that neutrinos, like the remaining charged fermions of the SM, are after all Dirac fermions. The Dirac approach to neutrino mass generation has also been widely explored in recent literature, and typically requires extra symmetries to forbid operators that lead to Majorana neutrino masses~\cite{Aranda:2013gga}. 
This is the case for the type-I~\cite{Ma:2014qra,Addazi:2016xuh,Bonilla:2017ekt} or type-II~\cite{Valle:2016kyz,Reig:2016ewy,Bonilla:2016zef}
tree-level Dirac seesaw mechanism.
Radiative neutrino at the one-loop level has also been discussed, e.g. within the context of the scotogenic framework ~\cite{Bonilla:2016diq,Bonilla:2018ynb,Leite:2020bnb,Leite:2020wjl} (for a comprehensive analysis see Ref.~\cite{CentellesChulia:2024iom}).

Another longstanding and profound puzzle in particle physics is the apparent conservation
of the strong charge-parity~(CP) symmetry in Quantum Chromodynamics~(QCD). 
Indeed, the phase $\bar{\theta}$ encoding CP violation must be tiny,  $\left|\overline{\theta}\right| < 10^{-10}$~~\cite{PhysRevD.92.092003,PhysRevLett.97.131801}, a fact that poses the so-called strong CP problem: why is CP significantly broken in weak interactions, while strong interactions seem to conserve it? One of the most popular proposed solutions to this problem is the Peccei-Quinn~(PQ) mechanism~\cite{PhysRevLett.38.1440,PhysRevD.16.1791}, where a new global, chiral U(1) symmetry, denoted $\text{U}(1)_{\text{PQ}}$, is postulated. 
The spontaneous breaking of $\text{U}(1)_{\text{PQ}}$ leads to a Nambu-Goldstone boson - the axion~\cite{PhysRevLett.40.223,PhysRevLett.40.279}. 
When the PQ symmetry is spontaneously broken, it generates a contribution of the same form as the $\overline{\theta}$-term. 
The phase $\overline{\theta}$ is promoted to the dynamical axion field, whose vacuum minimum effectively sets $\overline{\theta} = 0$, solving the strong CP problem.
 This symmetry is also explicitly broken by the QCD anomaly, giving a small mass to the axion. 
 
There are two main invisible axion frameworks, originally proposed
by Kim-Shifman-Vainshtein-Zakharov~(KSVZ)~\cite{Kim:1979if,Shifman:1979if}
and also the Dine-Fischler-Srednicki-Zhitnitsky~(DFSZ) axion~\cite{Zhitnitsky:1980tq,Dine:1981rt}.
Both scenarios rely on the spontaneous breaking of a global PQ symmetry via a complex scalar singlet, giving rise to the axion.
In the former, exotic fermions are introduced, and charged under the PQ symmetry, while in the latter, the SM quarks are charged under the PQ symmetry, and no exotic quarks are assumed.
Axions, initially proposed to solve the strong CP problem, are also excellent dark matter~(DM) candidates~\cite{Preskill:1982cy,Abbott:1982af,Dine:1982ah}, thus offering another solution to a major challenge of particle physics within the SM, namely the absence of a viable DM candidate. 

Reconciling the axion paradigm
with an appealing mechanism of neutrino mass generation showcases an interesting idea. 
With this in mind, we consider the possibility of embedding axions in Dirac neutrino mass generation schemes, as in Refs.~\cite{Gu:2016hxh,Peinado:2019mrn,Dias:2020kbj,delaVega:2020jcp,Berbig:2022pye}. 
We do this within a novel realization of the recent color-mediated neutrino mass framework proposed in~\cite{Batra:2023erw}. 
In this scenario, the colored particles necessary to solve the strong CP problem are directly responsible for mediating non-zero neutrino masses radiatively, with the axion as the unique DM candidate. Inspired by these ideas, we propose a KSVZ-type axion framework where the colored fermions, required to solve the strong CP problem, together with scalar leptoquarks, mediate PQ-preserving Dirac neutrino masses at the one-loop level. In Refs.~\cite{Gu:2016hxh,Peinado:2019mrn,Dias:2020kbj,delaVega:2020jcp,Berbig:2022pye,Carenza:2019pxu}, the PQ symmetry ensures the \emph{Diracness} of neutrinos but neutrino mass generation proceeds e.g. via a Dirac seesaw mechanism~\cite{Peinado:2019mrn}. In contrast, our model unifies the origin of Dirac neutrino masses and the axion solution, so that colored fermions mediate neutrino mass generation through a loop diagram. In Sec.~\ref{sec:Framework}, we present the various models, showing how they address the strong CP problem~(Sec.~\ref{sec:strongCP}) and neutrino mass generation (Sec.~\ref{sec:Diracneutrinos}). This is arranged in such a way as to allow for the heavy colored states to decay into ordinary matter. We also briefly discuss how the QCD axion can account for cosmic DM abundance within a post-inflationary scenario (in Sec.~\ref{sec:DM}).
Our models can be experimentally distinguished through their predictions for the axion-to-photon coupling as well as flavor-violating axion interactions, as analysed in Sec.~\ref{sec:axionphoton} and~\ref{sec:axionflavorviolating}, respectively. Finally, our concluding remarks are made in Sec.~\ref{sec:concl}. Details regarding the scalar sector and heavy-light quark mixing are presented in the appendices.

\section{Framework}
\label{sec:Framework}

In Tables~\ref{tab:general} and~\ref{tab:postcharges}, we present the particle content and transformation properties of the fields under the SM and PQ symmetries in our color-mediated Dirac neutrino mass models. 
\begin{table}[t!]
\renewcommand*{\arraystretch}{1.5}
	\centering
	\begin{tabular}{| K{4cm} | K{3cm} | K{5.5cm} | K{3cm} |}
		\hline 
&Fields&\SM&  U$(1)_{\text{PQ}}$   \\
		\hline \hline
		\multirow{3}{*}{Leptons} 
&$\ell_L$&($\mathbf{1},\mathbf{2}, {-1/2}$)& $1/6$   \\
&$e_R$&($\mathbf{1},\mathbf{1}, {-1}$)& {$1/6$}    \\
&$\nu_R$&($\mathbf{1},\mathbf{1}, {0}$)& {$4/6$}  \\ \hline
Vector-like quarks &$\Psi_{1,2 L};\Psi_{1,2 R}$&($\mathbf{3},\mathbf{n}_\Psi, y_\Psi$)& {$\mathcal{Q}_{\text{PQ}}; \mathcal{Q}_{\text{PQ}}-1/2$}  \\ 
\hline \hline
\multirow{2}{*}{Scalars}  &$\Phi$&($\mathbf{1},\mathbf{2}, 1/2$)& {$0$}  \\
&$\sigma$&($\mathbf{1},\mathbf{1}, 0$)& {$1/2$}  \\ \hline
\multirow{2}{*}{Scalar Leptoquarks} &$\eta$&($\mathbf{3},\mathbf{n}_\eta \equiv \mathbf{n}_\Psi \pm \mathbf{1}, y_\Psi+1/2$)& {$\mathcal{Q}_{\text{PQ}}-4/6$}   \\
&$\chi$&($\mathbf{3},\mathbf{n}_\Psi, y_\Psi$)& {$\mathcal{Q}_{\text{PQ}}-4/6$}   \\		
\hline
	\end{tabular}
	\caption{\footnotesize Field content and transformation properties under \SM and U$(1)_\text{PQ}$. Here, $\mathbf{n}_{\Psi,\eta}$ indicate the SU(2)$_L$ representations of $\Psi$ and $\eta$, while $y_\Psi$ is the hypercharge of $\Psi$ and $\mathcal{Q}_{\text{PQ}}$ is a PQ charge parameter. Specific assignments are given in Table~\ref{tab:postcharges}.}
	\label{tab:general} 
\end{table}
\begin{table}[t!]
\renewcommand*{\arraystretch}{1.5}
	\centering
	\begin{tabular}{| K{1cm} | K{1cm} | K{1cm} | K{1cm} | K{6cm} | K{4cm} |}
		\hline 
$\mathbf{n}_\Psi$ & $y_\Psi$ & $\mathcal{Q}_{\text{PQ}}$ &$\mathbf{n}_\eta$ & Heavy-light quark mixing terms & Other decay terms  \\
		\hline \hline
		\multirow{4}{*}{$\mathbf{1}$} 
&\multirow{2}{*}{$-1/3$} &$1/2$& \multirow{4}{*}{$\mathbf{2}$} & $\overline{q_L} \Phi \Psi_R, \overline{\Psi_L} \sigma d_R $& $\overline{\ell_L} \tilde{\eta} d_R  $ \\
& &$0$& & $\overline{\Psi_L} d_R$ & $ \overline{q_L} \eta \nu_R $ \\
\cline{2-3} \cline{5-6} 
&\multirow{2}{*}{$2/3$} &$1/2$& & $\overline{q_L} \tilde{\Phi} \Psi_R, \overline{\Psi_L} \sigma u_R$& $\overline{\ell_L} \tilde{\eta} u_R, \overline{q_L} \eta e_R $ \\
& &$0$& & $\overline{\Psi_L} u_R$ & - \\
        \hline \hline
        \multirow{4}{*}{$\mathbf{2}$} 
&\multirow{2}{*}{$1/6$} &$1/2$& \multirow{4}{*}{$\mathbf{1},\mathbf{3}$} & $\overline{q_L} \Psi_R$& $\overline{\ell_L} \chi^\ast d_R$ \\
& &$0$& & $\overline{q_L} \sigma \Psi_R,\overline{\Psi_L} \Phi d_R, \overline{\Psi_L} \tilde{\Phi} u_R$ & - \\
\cline{2-3} \cline{5-6}
&$-5/6$ & \multirow{2}{*}{$0$} & & $\overline{\Psi_L} \tilde{\Phi} d_R$& - \\
\cline{2-2} \cline{5-6}
&$7/6$ & & & $\overline{\Psi_L} \Phi u_R$ & - \\
        \hline \hline
        \multirow{4}{*}{$\mathbf{3}$} 
&\multirow{2}{*}{$-1/3$} & \multirow{4}{*}{$0$} &$\mathbf{2}$ & \multirow{2}{*}{$\overline{q_L} \Phi \Psi_R$}& $\overline{\ell_L} \tilde{\eta} d_R  $ \\
& & & $\mathbf{4}$ & & - \\
\cline{2-2} \cline{4-6}
&\multirow{2}{*}{$2/3$} & & $\mathbf{2}$ &\multirow{2}{*}{$\overline{q_L} \tilde{\Phi} \Psi_R$}& $\overline{\ell_L} \tilde{\eta} u_R, \overline{q_L} \eta e_R $ \\
& & & $\mathbf{4}$ & & - \\
        \hline
	\end{tabular}
	\caption{\footnotesize Charge assignments for the new colored fields given in terms of the $\mathbf{n}_\Psi$, $y_\Psi$, $\mathcal{Q}_{\text{PQ}}$ and $\mathbf{n}_\eta$ parameters (column 1-4) of Table~\ref{tab:general}. Allowed VLQ and leptoquark Yukawa interactions leading to exotic particle decays are given in columns 5 and 6, respectively. See text for details.}
\label{tab:postcharges} 
\end{table}

In order to implement massive Dirac neutrinos we add three right-handed~(RH) neutrino fields $\nu_R$ to the SM particle content. In addition, to address the strong CP problem, we add chiral vector-like quarks~(VLQ) $\Psi_{L,R}$ charged under a global PQ symmetry U(1)$_{\text{PQ}}$, which is spontaneously broken by a complex scalar singlet $\sigma$. 
Color-mediated radiative Dirac neutrino masses are generated by adding to the scalar sector two leptoquark multiplets $\eta$ and $\chi$. Before proceeding, a few important comments on our construction are in order. Namely: 
\begin{itemize}
    \item The PQ symmetry ensures the Dirac nature of neutrinos and forbids bare Majorana neutrino mass terms. The PQ-breaking scalar $\sigma$ may lead to effective Majorana neutrino mass generation operators, written generically as,
    \begin{align} 
    \mathcal{L}_{\text{Maj.}} = \frac{\boldsymbol{\kappa}_{\text{Maj.}}}{2 \Lambda^{n + n^\prime + 1}} \; (\bar{\ell_L^c} \tilde{\Phi}^\ast) (\tilde{\Phi}^\dagger \ell_L) \; \sigma^n \sigma^{\ast n^\prime} + \text{H.c.} \; ,
    \label{eq:MajoranaOperators}
    \end{align}
    with $\ell_L$ and $\Phi$ being the SM lepton and Higgs doublet, respectively, and $\tilde{\Phi}=i \tau_2 \Phi$, with $\tau_2$ being the complex Pauli matrix.
   Here $\boldsymbol{\kappa}_{\text{Maj.}}$ is a  dimensionless Majorana-type coupling matrix and $\Lambda$ parameterizes some high-energy scale (the case $n=n^\prime=0$ corresponds to the well-known Weinberg operator~\cite{Weinberg:1980wa}).  
   Notice, however, that the lepton PQ assignments in Table~\ref{tab:general} were specifically chosen such that there is a residual $\mathcal{Z}_3$ symmetry under which $(\ell_L,e_R,\nu_R) \to \omega (\ell_L,e_R,\nu_R)$ and $(\eta,\chi) \to \omega^2 (\eta,\chi)$. This forbids all the above Majorana type operators. In contrast, the effective Dirac neutrino mass operators 
    \begin{equation}
    \mathcal{L}_{\text{Dirac}} = \frac{\boldsymbol{\kappa}_{\text{Dirac}}}{\Lambda^{n+n^\prime}} \; (\bar{\ell_L} \tilde{\Phi} \nu_R) \; \sigma^n \sigma^{\ast n^\prime} + \text{H.c.} \; ,
    \label{eq:DiracOperators}
    \end{equation}
    are allowed by the $\mathcal{Z}_3$ symmetry with the lowest dimension operator being the dimension 5 $(\bar{\ell_L} \tilde{\Phi} \nu_R) \; \sigma^\ast$, i.e. for $n=0,n^\prime=1$.  
    This method of residual $\mathcal{Z}_N$ symmetries forbidding Majorana neutrino masses has been previously employed in the context of a $\text{U}(1)_{B-L}$ symmetry in Ref.~\cite{Bonilla:2018ynb}. 
    The tree-level Yukawa interaction $\bar{\ell_L} \tilde{\Phi} \nu_R$ is forbidden by the PQ symmetry, which is necessary for radiative Dirac neutrino mass generation. 
    In the above, $\boldsymbol{\kappa}_{\text{Dirac}}$ is a dimensionless Dirac-type neutrino coupling matrix. In Sec.~\ref{sec:Diracneutrinos} we will present the explicit one-loop realization of this dimension-5 operator. 
    \item All our models feature mixing between the VLQ $\Psi$ and the ordinary SM quarks ($q_L$ \rd{is} the quark doublet, $u_R,d_R$ up and down-type quark singlets).  
    Namely, seven possible VLQ representations mix with the SM quarks at tree-level~\cite{delAguila:2000aa,delAguila:2000rc,Aguilar-Saavedra:2013qpa,DiLuzio:2016sbl}, as shown in Table~\ref{tab:postcharges} (see fifth column). 
    In Appendix~\ref{sec:quarksector}, we discuss heavy-light quark mixing properties in detail for each model. 
    We will focus on these seven color-mediated Dirac neutrino mass models since they allow the new exotic colored particles $\Psi,\eta,\chi$ to decay into ordinary matter. 
    Otherwise one would have stable baryonic and charged relics which may pose cosmological difficulties~\cite{Perl:2001xi,Perl:2009zz,Burdin:2014xma,Hertzberg:2016jie,Mack:2007xj}.  
    Thus, our framework favors a post-inflationary axion DM cosmology, as discussed in Sec.~\ref{sec:DM}~\footnote{ A pre-inflationary scenario can also be envisaged and the charges of the new fields under the SM gauge group can be more general, as shown in Ref.~\cite{Batra:2023erw}. Axion DM cosmology is discussed in Sec.~\ref{sec:DM}.}.

    \item There is no proton decay in our models. The scalar leptoquarks $\eta$ and $\chi$ lead to the lepton-quark interaction terms given in the last column of Table~\ref{tab:postcharges}. 
    However, an accidental baryon number symmetry under which the SM quark fields and the new colored ones are equally charged forbids dangerous proton decay operators such as the dimension 6 $duue$, $qque$, $q\ell d u$, $qqql$, etc, with $e_R$ being the SM charged-lepton singlet fields. 
    This distinctive feature of our construction ensures that, besides protecting the \emph{Diracness} of neutrinos, the same PQ symmetry necessary to solve the strong CP problem, also ensures the stability of the proton.

    \item It must be noted that our setup relies on a global PQ symmetry which is susceptible to Planck-scale violations leading to the well-known ``PQ quality problem"~\cite{Georgi:1981pu,Dine:1986bg,Barr:1992qq,Kamionkowski:1992mf,Holman:1992us,Ghigna:1992iv}. This is analogous to issues in neutrino mass models based on global U$(1)_L$/U$(1)_{B-L}$ symmetries, which give rise to the Majoron -- see e.g. Ref.~\cite{Berezinsky:1993fm} and references therein. Addressing this theoretical challenge lies beyond the scope of our present work. Nevertheless, it has been shown that high-quality PQ symmetries can emerge as accidental remnants of discrete gauge symmetries~\cite{Chun:1992bn,BasteroGil:1997vn,Babu:2002ic,Dias:2002hz, Harigaya:2013vja}, Abelian gauge symmetries~\cite{Fukuda:2017ylt,Duerr:2017amf,Bonnefoy:2018ibr}, or non-Abelian gauge symmetries~\cite{Randall:1992ut,DiLuzio:2017tjx,Lillard:2018fdt, Lee:2018yak}, being promising avenues to explore in future studies.
    
\end{itemize}

We now explain how our unified framework solves the strong CP problem, neutrino mass generation as well as the cosmological DM problem.

\subsection{Strong CP problem}
\label{sec:strongCP}

The PQ scalar field can be written as $\sigma = (v_\sigma + \rho) \exp(i a /v_\sigma)/\sqrt{2}$, where $a$ is the axion and $\rho$ the radial mode. Once~$\sigma$ develops a non-zero vacuum expectation value~(VEV), the PQ symmetry is spontaneously broken at a scale $f_{\text{PQ}} = \left< \sigma \right>$. The axion decay constant~$f_a$ is related to  $f_{\text{PQ}}$ as follows,
\begin{equation}
    f_a = \frac{f_{\text{PQ}}}{N} = \frac{v_{\sigma}}{\sqrt{2} N} \; ,
    \label{eq:axionfa}
\end{equation}
with $N$ being the color anomaly factor. The up-to-date next-to-leading order (NLO) calculation of the axion mass yields~\cite{GrillidiCortona:2015jxo},
\begin{equation}
    m_a = 5.70(7) \left(\frac{10^{12} \text{GeV}}{f_a}\right) \mu \text{eV} \; .
    \label{eq:axionmass}
\end{equation}
This relation between $m_a$ and $f_a$ is a model-independent prediction of the QCD axion when the PQ symmetry is broken only by the QCD instantons generating the $G \tilde{G}$ term. In order to have a viable axion solution to the strong CP problem, the color anomaly factor $N$ must be non-vanishing, so that the axion couples to gluons via $G \tilde{G}$. For the class of models given in Table~\ref{tab:general}, we have:
\begin{equation}
N = 2 \sum_f \left(\omega_L^f - \omega_R^f\right) T(R_f) \; ,
\label{eq:Nmodel}
\end{equation}
where $f$ is a fermion $\Psi_f$ with PQ charge $\omega_f$ and representation $R_f$ under QCD with associated Dynkin index $T(R_f)$. The PQ charge assignments given in Table~\ref{tab:general} lead to $N=1,2,3$ for the cases where the VLQ $\Psi$ are iso-singlet, doublets, and triplets, respectively.

\subsection{Radiative Dirac neutrino masses}
\label{sec:Diracneutrinos} 
 
\begin{figure}[t!]
        \centering
        \includegraphics[scale=0.9]{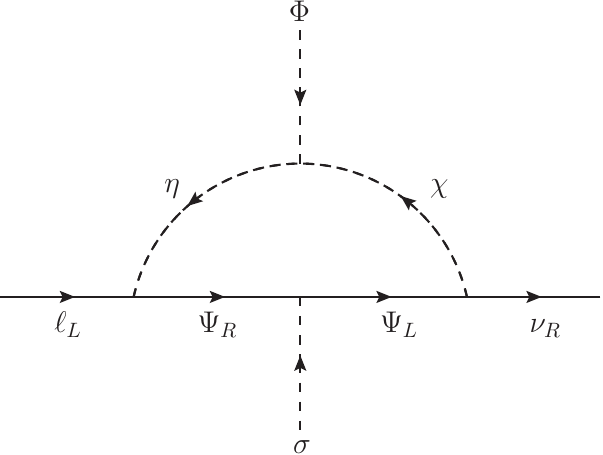}
        \caption{One-loop color-mediated Dirac neutrino masses (see Tables~\ref{tab:general} and~\ref{tab:postcharges}).}
    \label{fig:neutrinoDirac1loopcolor}
    \end{figure}

The Yukawa interactions responsible for neutrino mass generation are:
\begin{equation}
    - \mathcal{L}_{\text{Yuk.}}^{\nu} = \mathbf{Y}_{\eta} \overline{\ell_L} \; \tilde{\eta} \Psi_R + \mathbf{Y}_{\chi} \overline{\Psi_L} \chi \nu_R + \mathbf{Y}_{\Psi} \overline{\Psi_L} \Psi_R \sigma + \text{H.c.} \; ,
    \label{eq:LYukgenYneq0}
\end{equation}
while the relevant scalar-potential term is (see Appendix~\ref{sec:scalarsector} for the most general scalar potential), 
\begin{align}
    V \supset \kappa \; (\eta^\dagger \Phi) \chi  + \text{H.c.} \; .
    \label{eq:VneutrinoYneq0}
\end{align}
The above terms trigger one-loop color-mediated Dirac neutrino masses via the diagram shown in Fig.~\ref{fig:neutrinoDirac1loopcolor}. 
    In order to determine the neutrino mass matrix we must define couplings in the mass-eigenstate basis. Besides the VLQ $\Psi$, necessary to solve the strong CP problem, neutrino masses involve the states $\zeta_{1,2}$ arising from the $\eta$-$\chi$ scalar leptoquark mixing encoded in the $\mathbf{R}$ matrix. This, in turn, is controlled by the $\kappa$ scalar-potential parameter and the colored scalar masses $m_{\zeta_{1.2}}$ -- see Eq.~\eqref{eq:rotationscalars}. The Yukawa matrices in Eq.~\eqref{eq:LYukgenYneq0} are given in the mass-eigenstate basis as
\begin{align}
(\tilde{\mathbf{Y}}_\eta)_{\alpha j} & = \sum_{k=1}^{2} (\mathbf{Y}_\eta)_{\alpha k} (\mathcal{V}_R^q)_{(3+k) j} \; , \; (\tilde{\mathbf{Y}}_\chi)_{i \beta} = \sum_{k=1}^{2} (\mathcal{V}_L^{q \ast})_{(3+k) i} (\mathbf{Y}_\chi)_{k \beta} \; ,
 \end{align}
where $\alpha, \beta$ are flavor indices, $j,k$ run from 1 to 5, with the first three indices referring to the SM quarks and the remaining two corresponding to the heavy VLQ. $\mathcal{V}_{L,R}^{q}$ describe the mixing between $\Psi$ and SM quarks as seen in Eqs.~\eqref{eq:diagdownfull} and~\eqref{eq:qtrans}, with $q$ representing the up/down sector, depending on the VLQ representation, see Appendix~\ref{sec:quarksector}. Without loss of generality, we assume charged-leptons are in their mass basis, and also that $\mathbf{M}_\Psi \equiv v_\sigma \mathbf{Y}_\Psi/\sqrt{2} = \text{diag}(M_{\Psi_1}, M_{\Psi_2})$. Namely, their masses are given by $\mathbf{\tilde{m}}=(m_{q_1},m_{q_2},m_{q_3},\tilde{M}_{\Psi_1},\tilde{M}_{\Psi_2})$ where $q_{1,2,3}$ denote the SM quarks. From the above, the neutrino mass matrix is given by~\cite{CentellesChulia:2024iom},
\begin{align}
(\mathbf{M}_\nu)_{\alpha \beta} & = \frac{N_c}{16 \pi^2 \sqrt{2}} \sum_{j, k= 1}^{5} (\tilde{\mathbf{Y}}_\eta)_{\alpha j} (\tilde{\mathbf{Y}}_\chi)_{j \beta} \; \mathbf{R}_{1 k} \mathbf{R}_{2 k} \; \tilde{m}_{j} \frac{m_{\zeta_k}^2}{\tilde{m}_{j}^2-m_{\zeta_k}^2} \ln\left(\frac{\tilde{m}_{j}^2}{m_{\zeta_ k}^2}\right) \; ,
\label{eq:neutrinomass}
\end{align}
where the color factor $N_c=3$.

From this, we can estimate the neutrino masses as
\begin{align}
(M_\nu)_{\alpha \beta} & \sim 0.1 \text{eV} \; \left(\frac{(\tilde{Y}_\eta)_{\alpha j} (\tilde{Y}_\chi)_{j \beta}}{10^{-2}}\right)  \left(\frac{\kappa}{ 10^2 \; \text{GeV}}\right) \; \left(\frac{\tilde{M}_{\Psi j}}{10^{12} \; \text{GeV}}\right) \; \left(\frac{10^{10} \; \text{GeV}}{m_{\zeta_k}}\right)^2 \; ,
\end{align}
where we took typical values for the VLQ and colored leptoquark masses which are given by the PQ scale $f_{\text{PQ}}=v_\sigma/\sqrt{2}$. As we will see in Sec.~\ref{sec:DM}, $f_a \sim \mathcal{O}(10^{12})$ GeV is the typical scale required for the axion particle to account for the observed DM abundance. 

Notice that the parameter $\kappa$ can be naturally small in the t' Hooft sense~\cite{tHooft:1979rat}, as in the limit $\kappa \to 0$, the theory exhibits a larger symmetry. 
It follows that the smallness of Dirac neutrino masses is naturally controlled by the smallness of the $\zeta_{1,2}$ mass splitting. Moreover, the minimal number of $\Psi$ species to successfully generate the two observed neutrino mass splittings and lepton mixing is 2, which we choose here for simplicity.
Hence, these minimal color-mediated Dirac neutrino mass models predict a massless light neutrino. However, adding an extra $\Psi$ species would result in three non-zero light neutrino masses. As already mentioned, the underlying PQ symmetry in our construction enforces the {\em Diracness} of neutrinos.

\subsection{Axion dark matter and cosmology}
\label{sec:DM}

The attractiveness of the axion is further increased since it provides an excellent cold DM~(CDM) candidate, solving yet another unresolved puzzle of the SM. 
Axions are naturally light, weakly coupled with ordinary matter, nearly stable, and can be non-thermally produced (cold axions) in the early Universe. 

Axion DM can arise from two distinct scenarios: the pre-inflationary and post-inflationary cases, which we briefly outline here. 
In the pre-inflationary scenario, the PQ symmetry is broken before inflation and is never restored. In this case, axion DM is generated entirely through the misalignment mechanism~\cite{Preskill:1982cy,Abbott:1982af,Dine:1982ah}. Before inflation, different regions of the Universe have varying values for the axion field. Inflation subsequently expands one of these regions into our observable Universe. This implies that the axion field is constant over the observable Universe $a(\vec{x},t) = a(t) = \theta_0 f_a$, where $|\theta_0| \in [0, \pi)$ denotes the initial misalignment angle. 
When the QCD phase transition occurs at temperature $T \sim \Lambda_{\text{QCD}}$, the axion field is driven from its initial misalignment towards the vacuum minimum which preserves CP, solving the strong CP problem. The relic axion abundance generated through the misalignment mechanism can be approximately written as~\cite{DiLuzio:2020wdo} 
\begin{equation}
\Omega_a h^2 \simeq  \Omega_\text{CDM} h^2 \frac{\theta_0^2}{2.15^2} \left(\frac{f_a}{2 \times 10^{11} \ \text{GeV}} \right)^{\frac{7}{6}} \; ,
\label{eq:relica}
\end{equation}
where the observed CDM relic abundance, obtained by Planck satellite data is $\Omega_{\text{CDM}} h^2 = 0.1200 \pm 0.0012$~\cite{Planck:2018vyg}. From the above, one sees that for $\theta_0 \sim \mathcal{O}(1)$ we need $f_a \sim 5 \times 10^{11} \ \text{GeV}$ for axions to account for the full CDM abundance. This parameter region is currently being probed by haloscope experiments as will be discussed in Sec.~\ref{sec:axionphoton} -- see Fig.~\ref{fig:gagg}.

In the post-inflationary scenario, the PQ symmetry is broken after inflation. This leads to an observable Universe divided into patches with different values of the axion field. The initial misalignment angle $\theta_0$ is no longer a free variable and by performing a statistical average one obtains $\left<\theta_0^2\right> \simeq 2.15^2$~\cite{DiLuzio:2020wdo}. 
Thus, if $\Omega_a h^2 = \Omega_{\text{CDM}} h^2$, a prediction for $f_a$ (or equivalently $m_a$) can be made [see Eq.~\eqref{eq:relica}]. 
Indeed, if only the misalignment mechanism is at play, the upper bound of $f_a \lesssim 2 \times 10^{11}$ GeV ensures that DM is not overproduced. 
However, the picture is much more complicated since topological defects will also contribute to the total axion relic abundance $\Omega_a h^2$~\cite{Bennett:1987vf,Levkov:2018kau,Gorghetto:2018myk,Buschmann:2019icd}. 
The resulting network of strings and domain walls~(DWs) is complex, leading to non-linear dynamics that must be tackled through numerical simulations. 
Knowing their contribution to the relic abundance is still an unresolved problem. This analysis is beyond the scope of this work. Note, however, that a way to avoid the cosmological domain wall problem would be to have $N_{\text{DW}} \equiv N = 1$ [see Eq.~\eqref{eq:Nmodel}], where $N_{\text{DW}}$ corresponds to the vacuum degeneracy resulting from the residual $\mathcal{Z}_{N}$ symmetry left unbroken from non-perturbative QCD instanton effects that anomalously break U(1)$_{\text{PQ}}$. 
This feature occurs for the models of Tables~\ref{tab:general} and~\ref{tab:postcharges} containing an isosinglet VLQ [see Eq.~\eqref{eq:Nmodel}]. 
However, for the cases with an isodoublet and isotriplet exotic quark, we have $N_{\text{DW}}=2$ and $N_{\text{DW}}=3$, respectively, unavoidably leading to DWs in the early Universe~\cite{Lazarides:2018aev}. 
Note that for $N_{\text{DW}}=1$, although there are no DWs, string networks can still form, with current numerical simulations predicting the axion decay constant $f_a$ to lie around ($5 \times 10^9 - 3 \times 10^{11}$) GeV, such that $\Omega_a h^2 = \Omega_{\text{CDM}} h^2$~\cite{Buschmann:2021sdq,Gorghetto:2020qws,Klaer:2017ond,Kawasaki:2014sqa}. 

In contrast, in the pre-inflationary scenario, since U(1)$_{\text{PQ}}$ breaking occurs before inflation all topological defects, strings, and DWs, will be washed away. 
It follows that, even in the iso-doublet and triplet models, one may introduce a bias term in the scalar potential which slightly breaks the residual $\mathcal{Z}_{N}$ symmetry. 
This would lift the vacuum degeneracy and render the DW-string network unstable~\cite{Sikivie:1982qv}. The subsequent decay of cosmic strings and DWs can lead to very interesting stochastic gravitational wave signatures~\cite{Roshan:2024qnv,Servant:2023mwt,Morais:2023ciz}. 

\section{Axion phenomenology}
\label{sec:axionphoton}

The experimental axion program encompasses a wide array of setups designed to search for this elusive particle, with helioscopes and haloscopes standing out as prominent examples. Alongside indirect astrophysical and cosmological observations, these methods collectively constrain the axion parameter space by probing its couplings to photons, electrons, and nucleons. For detailed reviews see Refs.~\cite{DiLuzio:2020wdo,Adams:2022pbo} and the references therein. We also refer the reader to the repository~\cite{AxionLimits} which collects up-to-date searches for axions from a broad range of experiments. We begin this section by examining the various predictions for axion-to-photon couplings. In Sec.~\ref{sec:axionflavorviolating} we identify a few models featuring sizable axion-to-quark flavor violating couplings which are also constrained by quark flavor processes~\cite{MartinCamalich:2020dfe,Alonso-Alvarez:2023wig}.

\subsection{Axion-to-photon coupling}
\label{sec:axionphoton}

The models of Tables~\ref{tab:general}  and~\ref{tab:postcharges} will provide a distinct prediction for the axion-to-photon coupling $g_{a \gamma \gamma}$. Using NLO chiral Lagrangian techniques, we have~\cite{GrillidiCortona:2015jxo} 
\begin{align}
g_{a \gamma \gamma} &= \frac{\alpha_e}{2 \pi f_a} \left[\frac{E}{N} - 1.92(4) \right] \; .
\label{eq:gagg}
\end{align}
The ratio $E/N$ is the model-dependent contribution, with $E$ being the electromagnetic anomaly factor, given by 
\begin{equation}
E = 2 \sum_f \left(\omega_L^f - \omega_R^f\right) q^2_f \; ,
\label{eq:E}
\end{equation}
where $q_f$ is the electric charge of a fermion $\Psi_f$ transforming under U(1)$_{\text{PQ}}$. 
\begin{figure}[!t]
    \centering
      \includegraphics[scale=0.55]{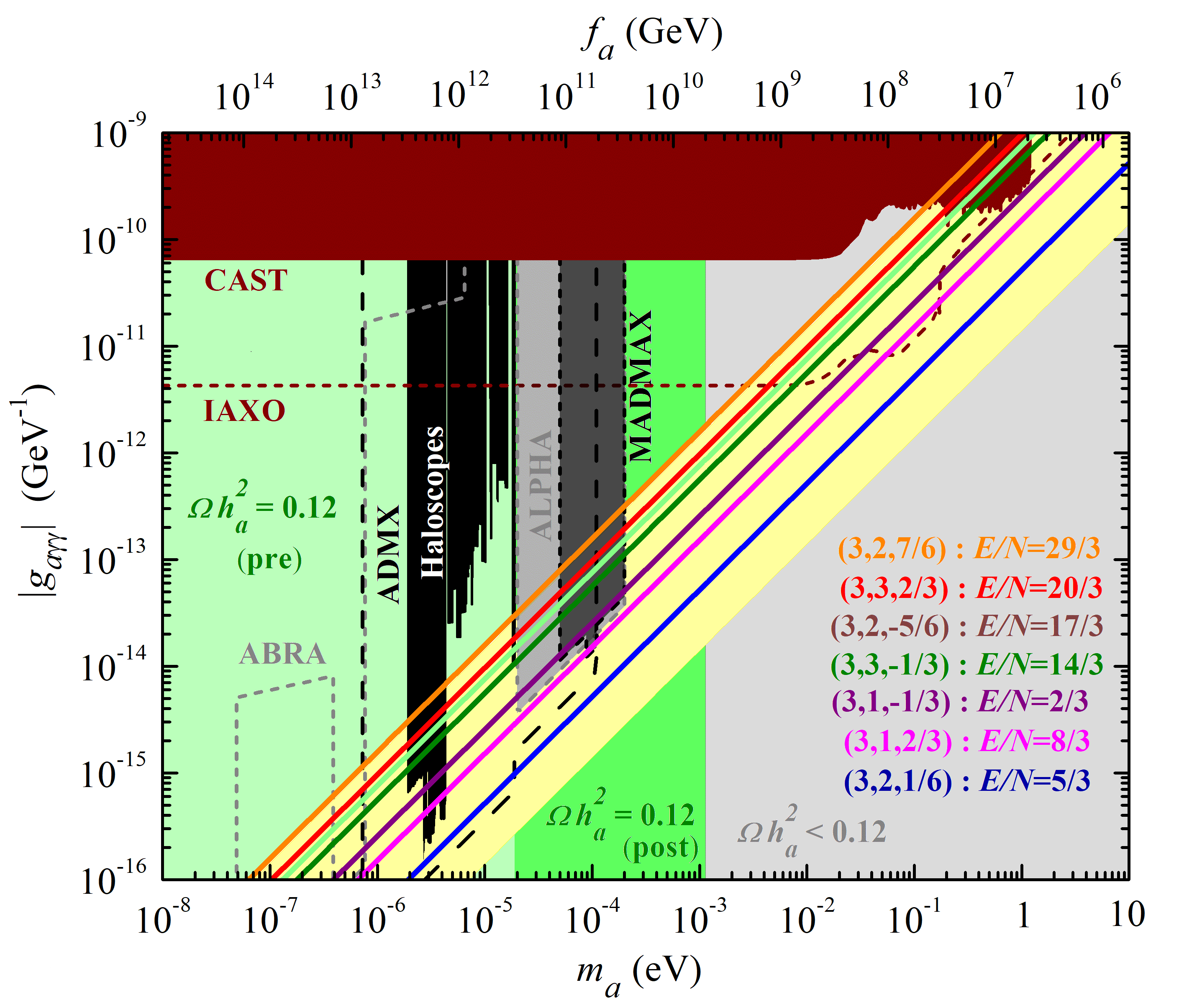}
    \caption{\footnotesize Axion-to-photon coupling $|g_{a \gamma \gamma}|$ versus the axion mass $m_a$ (bottom axis) and decay constant $f_a$ (top axis). The colored solid lines give the $E/N$ values of various Dirac color-mediated neutrino mass models (see Tables~\ref{tab:general} and~\ref{tab:postcharges}). The yellow shaded region encloses the usual QCD axion window~\cite{DiLuzio:2016sbl}. Current constraints from the helioscope experiment like CAST~\cite{CAST:2017uph} exclude the bordeau-shaded region, while haloscope experiments like ADMX~\cite{ADMX:2018gho,ADMX:2019uok,ADMX:2021nhd}, RBF~\cite{DePanfilis:1987dk}, CAPP~\cite{CAPP:2020utb} and HAYSTAC~\cite{HAYSTAC:2020kwv}, rule out the black region. The projected sensitivities of IAXO~\cite{Shilon_2013}, ADMX~\cite{Stern:2016bbw} and MADMAX~\cite{Beurthey:2020yuq} are indicated by the dashed bordeau, black contours and dark-gray shaded region, respectively. While the projected sensitivities of ABRACADABRA~\cite{Ouellet:2018beu} and ALPHA~\cite{ALPHA:2022rxj} are shown by the dashed gray contours and light-gray shaded region, respectively. In the gray band, axion DM is under-abundant. In the middle green band~$\Omega h^2_a = 0.12$ in the post-inflationary scenario for $N_{\text{DW}}=1$ ~\cite{Buschmann:2021sdq,Gorghetto:2020qws,Klaer:2017ond,Kawasaki:2014sqa}, while in the left green band $\Omega h^2_a = 0.12$ is still allowed in the pre-inflationary case.}
    \label{fig:gagg}
\end{figure}
In Fig.~\ref{fig:gagg}, we display $|g_{a\gamma \gamma}|$ in terms of the axion mass~$m_a$ (bottom axis) and decay constant~$f_a$ (top axis), showing current bounds and future sensitivities from helioscope and haloscope experiments. 
The colored oblique lines indicate the $|g_{a\gamma \gamma}|$ predictions for the scenarios presented in Tables~\ref{tab:general}  and~\ref{tab:postcharges}. Note that, the isosinglet up- (pink line) and down-type (purple line) VLQ predictions are identical to the popular DFSZ-I and DFSZ-II schemes~\cite{Zhitnitsky:1980tq,Dine:1981rt}, respectively. 
The upper and lower limits of $|g_{a\gamma \gamma}|$ correspond to the exotic fermion representations $\Psi \sim (\mathbf{3},\mathbf{2},7/6)$ (orange line) and $\Psi \sim (\mathbf{3},\mathbf{2},1/6)$ (dark blue line), respectively. 
Helioscope experiments aim at detecting the solar axion flux, with the CERN Solar Axion Experiment~(CAST)~\cite{CAST:2017uph} excluding $|g_{a \gamma \gamma}| \geq 0.66 \times 10^{-10} \ \text{GeV}^{-1}$ for masses $m_a \leq 20$ meV (bordeau-shaded region). 
The forthcoming International Axion Observatory~(IAXO)~\cite{Shilon_2013} is expected to probe the axion-to-photon coupling $g_{a \gamma \gamma}$ down to $(10^{-12}-10^{-11}) \ \text{GeV}^{-1}$ for $m_a \sim 0.1$ eV (bordeau-dashed contour), which scrutinizes all our models except for $\Psi \sim (\mathbf{3},\mathbf{2},1/6)$ (dark blue line). 
The most promising experiments are the haloscopes, which aim at detecting non-relativistic axions with the assumption that they make up the totality of the observed DM in the Universe (see Sec.~\ref{sec:DM}). 
These experiments include ADMX~\cite{ADMX:2018gho,ADMX:2019uok,ADMX:2021nhd}, RBF~\cite{DePanfilis:1987dk}, CAPP~\cite{CAPP:2020utb} and HAYSTAC~\cite{HAYSTAC:2020kwv}, which exclude the black shaded region. 

As discussed in Sec.~\ref{sec:DM}, the QCD axion can account for the observed DM relic abundance via the misalignment mechanism. In the post-inflationary scenario, for $N_{\text{DW}}=1$, numerical simulations of axion strings restrict the axion decay constant to the range $f_a \in [5 \times 10^9, 3 \times 10^{11}]$ GeV~\cite{Adams:2022pbo, Buschmann:2021sdq, Gorghetto:2020qws, Klaer:2017ond, Kawasaki:2014sqa} (middle green band). In contrast, the grey-shaded region on the right indicates under-abundant axion DM while in the dark green region $\Omega h^2_a = 0.12$ is still possible for the pre-inflationary scenario.
Out of all haloscopes, the most impressive is the Axion Dark Matter eXperiment~(ADMX). In fact ADMX has already excluded a considerable part of the parameter space of our scenarios. A notable exception occurs for $\Psi \sim (\mathbf{3},\mathbf{2},1/6)$ (dark blue line), for masses $m_a \sim 3 \ \mu$eV corresponding to $f_a \sim 10^{12}$ GeV. Future ADMX generations (dash-dotted black contour) are expected to probe all our models for masses $1 \ \mu \text{eV} \lesssim m_a \lesssim 20 \ \mu$eV or equivalently for scales $5 \times 10^{11} \ \text{GeV} \lesssim f_a \lesssim 10^{13}$ GeV. ALPHA~\cite{Lawson:2019brd,Wooten:2022vpj,ALPHA:2022rxj} (plasma haloscope) and MADMAX~\cite{Beurthey:2020yuq} are projected to probe the mass region $[20,200] \ \mu \text{eV}$ (light-gray region) and $[50,200] \ \mu \text{eV}$ (dark-gray region), respectively, covering all our frameworks except for $\Psi \sim (\mathbf{3},\mathbf{1},2/3)$ (pink line) and $\Psi \sim (\mathbf{3},\mathbf{2},1/6)$ (dark blue line). Lastly, ABRACADABRA~\cite{Ouellet:2018beu}, is expected to probe the mass region below $\sim 1 \mu \text{eV}$ and reach the yellow QCD axion band (dash gray contour), being an important complementary probe of axion DM compared to ``traditional" haloscope setups like ADMX. These experiments provide a sensitive probe of our models, complementary to other new physics searches at colliders or rare processes in the flavor sector.

\subsection{Flavor-violating axion couplings}
\label{sec:axionflavorviolating}

As in the conventional KSVZ scenarios~\cite{Kim:1979if,Shifman:1979if}, in our framework the SM quarks are not charged under the PQ-symmetry, resulting in no direct model-dependent couplings with the axion. Nonetheless, all color-mediated Dirac neutrino mass models in Tables~\ref{tab:general} and~\ref{tab:postcharges} exhibit mixing between the heavy VLQ $\Psi$ and the ordinary SM quarks. This will induce flavor-violating axion-quark couplings. 
The heavy-light quark mixing properties of the various models are studied in detail in Appendix~\ref{sec:quarksector}. They are described in terms of the $3\times 2$ $\mathbf{\Theta}_{X}^{q}$ mixing matrices of Eqs.~\eqref{eq:HLmixingrefdown},~\eqref{eq:HLmixing2SM},~\eqref{eq:HLmixing3down} and~\eqref{eq:HLmixing3up}, with $X=L,R$ and $q=d,u$ depending on the specific model. 
Without loss of generality, we can take $\mathcal{Q}_\text{PQ} = 0$ for all models, except for $\Psi \sim (\mathbf{3}, \mathbf{2}, 1/6)$ where we take $\mathcal{Q}_\text{PQ} = 1/2$ (see Table~\ref{tab:postcharges}). 
 These have the least number of mixing terms and hence simpler quark mass matrices. The results, summarized in
 Table~\ref{tab:axionflavorviolating}, can be easily generalized to the other cases.
\begin{table}[t!]
\renewcommand*{\arraystretch}{1.5}
	\centering
	\begin{tabular}{| K{0.5cm} | K{1cm} | K{1cm} | K{5.5cm} | K{7.5cm} |}
		\hline 
$\mathbf{n}_\Psi$ & $y_\Psi$ & $\mathcal{Q}_{\text{PQ}}$ & Heavy-light quark mixing terms & $\mathbf{\Theta}_{X}^{q}$ mixing parameter \\
		\hline \hline
\multirow{2}{*}{$\mathbf{1}$} & $-1/3$ &\multirow{2}{*}{$0$}& $ \mathbf{M}_{\Psi d} \overline{\Psi_L} d_R$ & $\boxed{\Theta_{R}^{d} \sim M_{\Psi d}/M_\psi} \ , \ \Theta_{L}^{d} \sim (v/M_\Psi) Y_d \Theta_{R}^{d}$ \\
& $2/3$ & & $ \mathbf{M}_{\Psi u} \overline{\Psi_L} u_R$ & $\boxed{\Theta_{R}^{u} \sim M_{\Psi u}/M_\psi} \ , \ \Theta_{L}^{u} \sim (v/M_\Psi) Y_u \Theta_{R}^{u}$ \\
        \hline \hline
        \multirow{3}{*}{$\mathbf{2}$} 
&$1/6$ &$1/2$& $\mathbf{M}_{q \Psi} \overline{q_L} \Psi_R$& $\boxed{\Theta_{L}^{d,u} \sim M_{q \Psi}/M_\psi} \ , \ \Theta_{R}^{d,u} \sim (v/M_\Psi) Y_{d,u} \Theta_{L}^{d,u}$ \\
\cline{2-5}
&$-5/6$ & \multirow{2}{*}{$0$} & $\mathbf{Y}_{\Psi d} \overline{\Psi_L} \tilde{\Phi} d_R$& $\Theta_{R}^{d} \sim (v/M_\psi) Y_{\Psi d} \ , \ \Theta_{L}^{d} \sim (v/M_\Psi) Y_d \Theta_{R}^{d}$ \\
&$7/6$ & & $ \mathbf{Y}_{\Psi u} \overline{\Psi_L} \Phi u_R$ & $\Theta_{R}^{u} \sim (v/M_\psi) Y_{\Psi u} \ , \  \Theta_{L}^{u} \sim (v/M_\Psi) Y_u \Theta_{R}^{u}$ \\
        \hline \hline
        \multirow{2}{*}{$\mathbf{3}$} 
&$-1/3$ & \multirow{2}{*}{$0$} & $ \mathbf{Y}_{q \Psi} \overline{q_L} \Phi \Psi_R$& $\Theta_{R}^{d} \sim (v/M_\psi) Y_{q \Psi} \ , \ \Theta_{L}^{d} \sim (v/M_\Psi) Y_d \Theta_{R}^{d}$ \\
&$2/3$ & &$\mathbf{Y}_{q \Psi} \overline{q_L} \tilde{\Phi} \Psi_R$& $\Theta_{R}^{u} \sim (v/M_\psi) Y_{q \Psi} \ , \ \Theta_{L}^{u} \sim (v/M_\Psi) Y_u \Theta_{R}^{u}$ \\
        \hline
	\end{tabular}
	\caption{\footnotesize Heavy-light quark mixing terms and $\mathbf{\Theta}_{X}^{q}$ parameter for the various models in Tables~\ref{tab:general} and~\ref{tab:postcharges} characterized by the VLQ $\Psi$ representation. We highlight via a box the heavy-light quark mixing parameters that can be sizable (see text for details).}
\label{tab:axionflavorviolating} 
\end{table}

Among the seven potential models, three exhibit significant heavy-light quark mixing, resulting in substantial axion-flavor violating couplings. Specifically, VLQ $\Psi$ representations $(\mathbf{3}, \mathbf{1}, -1/3)$, $(\mathbf{3}, \mathbf{1}, 2/3)$ and $(\mathbf{3}, \mathbf{2}, 1/6)$, are the only ones with sizable heavy-light quark mixing, in the RH-down-quark, RH-up-quark and left-handed-(LH)-up- and down-quark sectors, respectively. This is highlighted in the table with boxes around the relevant sizable $\mathbf{\Theta}_{X}^{q}$ parameters. For the isosinglet model with $\Psi \sim (\mathbf{3}, \mathbf{1}, -1/3)$, we have a sizable $\mathbf{\Theta}_R^d$. In fact, since $\mathbf{M}_{\Psi d}$ is an arbitrary bare mass term we can take its scale $M_{\Psi d}$ close to the $M_\Psi \sim \mathcal{O}(f_{\text{PQ}})$ scale, implying a sizable $\Theta_R^d \sim M_{\Psi d}/M_\Psi$. In contrast, $\mathbf{\Theta}_L^d \sim \mathcal{O}(v/f_{\text{PQ}}) \mathbf{\Theta}_R^d \sim \mathcal{O}(10^{-10}) \mathbf{\Theta}_R^d$, is completely negligible. The same holds for $\Psi \sim (\mathbf{3}, \mathbf{1}, 2/3)$, with the substitution $d \rightarrow u$. On the other hand, the isodoublet case with $\Psi \sim (\mathbf{3}, \mathbf{2}, 1/6)$ will feature sizable heavy-light mixing $\mathbf{\Theta}_L^{d,u}$ with $\mathbf{\Theta}_R^{d,u}$ being completely negligible. As for the rest of the models, the heavy-light quark mixing is not controlled by an arbitrary bare mass parameter, but instead by a Yukawa coupling with the Higgs doublet, resulting in further $v/f_{\text{PQ}}\sim \mathcal{O}(10^{-10})$ suppression making the mixing negligible. Sizable mixing translates into sizable flavor-violating quark-axion couplings, as discussed below.

The heavy-light quark mixing induces flavor-changing neutral currents in the quark sector, involving the SM $Z$ boson and Higgs field $h$, as well as flavor-changing couplings of the axion field $a$. The VLQ $\Psi$ are chirally charged under the PQ symmetry with their LH and RH PQ charge difference matching $\mathcal{Q}_L - \mathcal{Q}_R = \mathcal{Q}_\sigma = 1/2$, where $\mathcal{Q}_\sigma$ is the PQ charge of $\sigma$. Depending on the specific model, either the LH or RH chiral component of $\Psi$ will be charged under PQ (see the $\mathcal{Q}_\text{PQ}$ charge values in Tables~\ref{tab:general} and~\ref{tab:postcharges}). Generically, $\Psi_{j X}$ transforms under the PQ symmetry as $\Psi_{j X}\to \exp(i \mathcal{Q}_X a /v_\sigma) \Psi_{j X}$. By rotating the axion field, we can remove it from the Yukawa Lagrangian and it will appear in the kinetic term for $\Psi_{j X}$~\footnote{Axion couplings can also be expressed in Cartesian   -- as opposed to polar -- form. The general method has been extensively used in describing Majoron couplings~\cite{Schechter:1981cv}.}. Since $\Psi_{j X}$ mixes with the SM quarks through $\mathbf{\Theta}_{X}^{q}$, one obtains the following flavor-violating axion interactions,
\begin{align}
    \mathcal{L}^a_{\rm FV} &= \frac{ \partial_\mu a}{ v_\sigma} \overline{q_{\alpha X}}\,\gamma^\mu \,\mathcal{Q}_X (\tilde{\mathbf{\Theta}}_X^q)_{\alpha \beta} \, q_{\beta X} = \frac{ \partial_\mu a}{f_a} \overline{q_{\alpha}}\,\gamma^\mu \,\left[ (c^q_V)_{\alpha \beta}  + (c^q_A)_{\alpha \beta} \gamma_5 \right]\, q_{\beta X} \;, \nonumber \\
    (c^q_V)_{\alpha \beta} & = (c^q_A)_{\alpha \beta} = c^q_{\alpha \beta} =\frac{1}{2} \mathcal{Q}_X (\tilde{\mathbf{\Theta}}_X^q)_{\alpha \beta} \; , \; \tilde{\mathbf{\Theta}}_{X}^{q} = {\mathbf{V}_{X}^{q}}^\dagger \mathbf{\Theta}_{X}^{q} {\mathbf{\Theta}_{X}^{q}}^\dagger \mathbf{V}_{X}^{q} \; ,
    \label{eq:FCNCa}
\end{align}
where $c_{V,A}^{\alpha \beta}$ represent the vector and axial couplings, which are Hermitian matrices in flavor space. 

In Refs.~\cite{MartinCamalich:2020dfe,Alonso-Alvarez:2023wig}, a detailed study of flavor-violating axion couplings to quarks resulted in an exhaustive list of constraints stemming from rare processes, Higgs physics, meson oscillations, among others have been discussed. These constraints were applied to the color-mediated Majorana neutrino mass framework~\cite{Batra:2023erw, Hati:2024ppg}. The goal in \cite{Hati:2024ppg} was to account for an enhanced $B^+\to K^+ + E_{\text{miss}}$ rate at Belle-II with a model featuring heavy-light mixing in the down-quark sector. The heavy-colored scalars mediating neutrino mass generation had masses around the TeV scale in order to explain the aforementioned flavor anomaly. Here, we consider more natural values for the masses of the heavy-colored scalars arising from the leptoquarks $\eta$ and $\chi$, and their mixing.  Since these scalars couple to $\sigma$ via a quartic interaction in the scalar potential (see Appendix~\ref{sec:scalarsector}), their masses should be close to the PQ breaking scale $f_{\text{PQ}} \sim \mathcal{O}(10^{12})$ GeV, if one assumes $\mathcal{O}(1)$ scalar potential couplings (see discussion in Sec.~\ref{sec:Diracneutrinos}). 

We also wish to mention that our Dirac color-mediated models, with $\Psi \sim (\mathbf{3}, \mathbf{1}, -1/3)$ and $\Psi \sim (\mathbf{3}, \mathbf{2}, 1/6)$, could in principle explain such quark flavor anomalies, thanks to the new colored scalars and sizable interactions in the down-quark sector controlled by $c_{V,A}^{\alpha \beta}$ (see boxed expressions in Table~\ref{tab:axionflavorviolating}). We do not pursue such a detailed analysis here, and will instead focus on axion flavor-violating couplings. Moreover, since a detailed study of the flavor violating axion to down-quark coupling was performed in Ref.~\cite{Hati:2024ppg}, here we wish to discuss the up-quark sector. 
\begin{figure}[!t]
    \centering
    \includegraphics[scale=0.175]{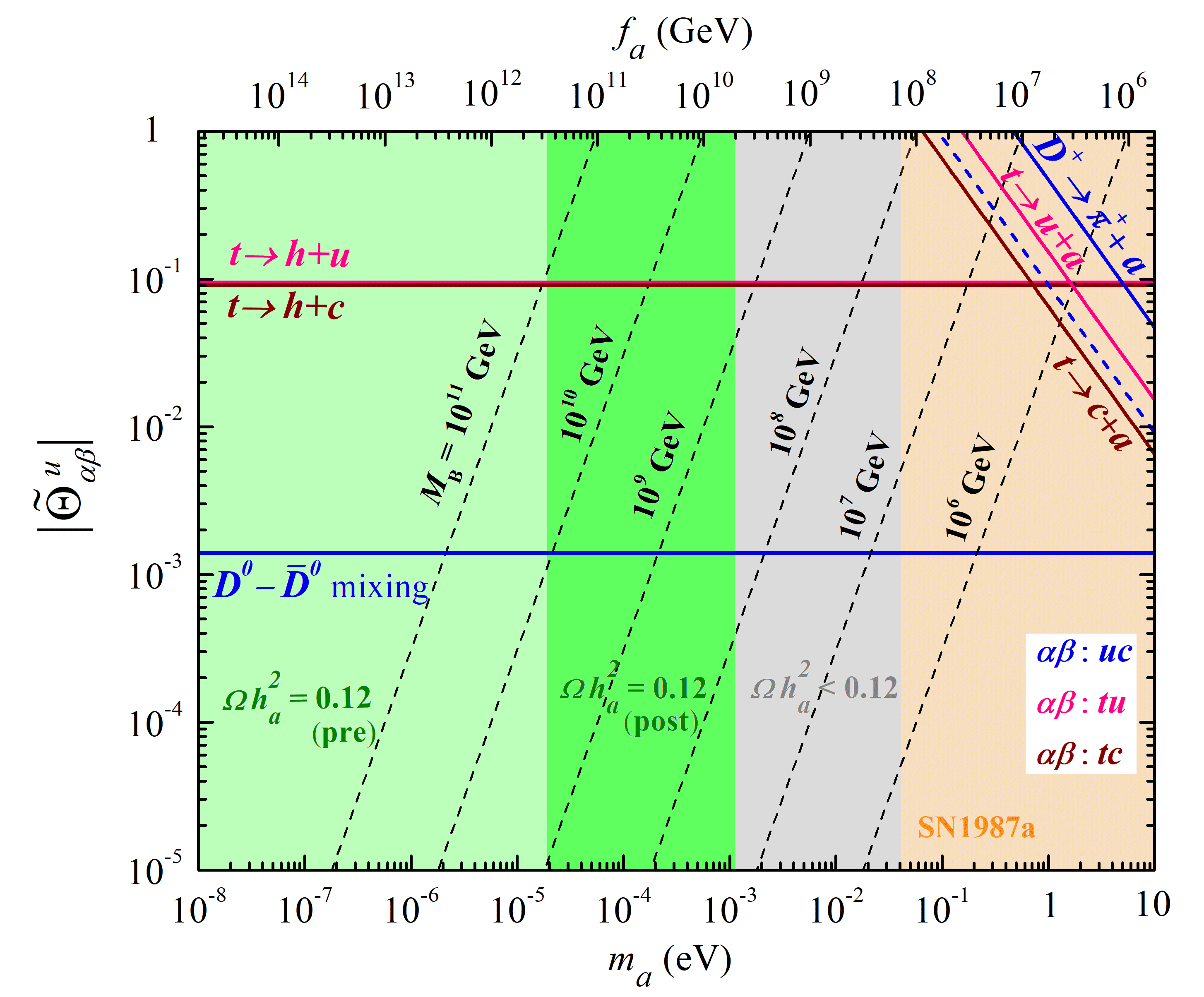}
    \caption{\footnotesize Flavor-violating axion to up-quark coupling, encoded in $|\tilde{\Theta}^{u}_{\alpha\beta}|$, versus $m_a$ (bottom axis) and $f_a$ (top axis). Constraints in blue (pink) [red], only apply to the case where $\alpha \beta \equiv uc$ (tu) [tc]~\cite{MartinCamalich:2020dfe,Alonso-Alvarez:2023wig} (see text for details).  Our benchmarks are indicated by black-dashed lines with $M_{\text{B}} \in [10^{6} - 10^{11}]$ GeV. The orange-shaded region is excluded by SN1987a~\cite{Carenza:2019pxu}. Shaded bands follow the same color code as in Fig.~\ref{fig:gagg}.}
    \label{fig:axionfvup}
\end{figure}
Among all the Dirac color-mediated neutrino mass models only two can have sizable axion-to-up-quark flavor violating couplings. Namely, the models with $\Psi \sim (\mathbf{3},\mathbf{1},2/3)$ and $\Psi \sim (\mathbf{3},\mathbf{1},1/6)$, with the mixing encoded in $\mathbf{\Theta}^{u}_R$ and $\mathbf{\Theta}^{u}_L$, respectively (see Table~\ref{tab:axionflavorviolating}). Following Eqs.~\eqref{eq:HLmixingrefdown}, replacing $d$ by $u$,  and ~\eqref{eq:HLmixing2SM}, we express the mixing parameter of Eq.~\eqref{eq:FCNCa} as, 
\begin{equation}
\mathbf{\tilde{\Theta}}^u_{\alpha \beta} =  \sum_{k=1}^2 \frac {(\mathbf{M}_{\text{B}})_{k\alpha} (\mathbf{M}_{\text{B}})_{k\beta} } {\mathbf{M}_{\Psi k}^2} = \sum_{k=1}^2 \frac {(\mathbf{M}_{\text{B}})_{k\alpha} (\mathbf{M}_{\text{B}})_{k\beta} } {(N f_a \mathbf{Y}_{\Psi k})^2} \,, 
\end{equation}
where the bare mass matrices $\mathbf{M}_{\Psi u}$ and $\mathbf{M}_{q \Psi}$ are given by $\mathbf{M}_{\text{B}}$. Note that, without loss of generality, we assume that $\mathbf{M}_{\Psi}$ is in the diagonal basis and, for simplicity, neglect CKM mixing. In Fig.~\ref{fig:axionfvup}, we present the axion-to-up-quark flavor-violating couplings via $|\tilde{\Theta}^{u}_{\alpha\beta}|$ [see Eq.~\eqref{eq:FCNCa}], in terms of the axion mass $m_a$ [or equivalently $f_a$ -- see Eq.~\eqref{eq:axionfa}]. We took six benchmarks for the bare mass parameter $M_{\text{B}}$, setting the scale between $10^{6} - 10^{11}$ GeV, taking $\mathbf{M}_{\Psi} = N f_a$. These cases are indicated by black-dashed lines. Moreover, the model-independent axion-nucleon couplings, induced by the axion-gluon interactions, are $C_{an} = -0.02$ and $C_{ap} = -0.47$, for the neutron and proton, respectively~\cite{Carenza:2019pxu,DiLuzio:2020wdo}. The main constraint on these couplings comes from supernova SN1987a, namely, $0.29\,  g_{ap}^2< 3.25 \times 10^{-18}$, where $g_{ap} \sim m_p C_{ap}/f_a$ with $m_p$ being the mass of proton. This translates into a lower bound on $f_a > 1.4 \times 10^8$ GeV~\cite{Carenza:2019pxu}, see the orange dashed region in the figure. As seen in Fig.~\ref{fig:gagg} the QCD axion can account for the observed DM relic abundance, in the post-inflationary scenario for $N_{\text{DW}}=1$, the axion decay constant is restricted to the range $f_a \in [5 \times 10^9, 3 \times 10^{11}]$ GeV~\cite{Adams:2022pbo, Buschmann:2021sdq, Gorghetto:2020qws, Klaer:2017ond, Kawasaki:2014sqa} (middle green band).

The constraints in blue (pink) [red], thoroughly studied and summarized in Refs.~\cite{MartinCamalich:2020dfe,Alonso-Alvarez:2023wig}, only apply to the case where $\alpha \beta \equiv uc$ ($tu$) [$tc$]. For the up-type mixing, the constraints imposed by rare quark flavor processes $D^{+}\to \pi^{+} \, + a$ are shown by solid and dashed blue lines to indicate current and future sensitivity. For top decays, $t\to u \, + a$  and $t\to c \, + a$ are presented using the pink and red lines, respectively. These constraints lie within the orange shaded region already excluded by SN1987a. However, since heavy-light quark mixing also leads to flavor changing neutral currents ($Z$ boson) and couplings (Higgs boson $h$), $D^0-\overline{D}^0$ meson mixing (horizontal blue line) and top decays into the Higgs $t\to h \, + u$ (horizontal pink line) and $t\to h \, + c$ (horizontal red line), can constrain $|\tilde{\Theta}^{u}_{\alpha\beta}|$ independently of the axion scale~$f_a$. In summary, taking into the account the DM requirement and all the above complementary constraints, the bare mass parameter of our models is restricted to lie within the range $M_B \in [10^8 - 10^{10}]$ GeV, where for $\alpha\beta = uc$ the mixing is bounded as $|\tilde{\Theta}^{u}_{\alpha\beta}|\lsim 10^{-3}$, while for $\alpha\beta = tu,tc$ we have $|\tilde{\Theta}^{u}_{tu,tc}|\lsim 10^{-1}$.
 
\section{Concluding remarks} 
\label{sec:concl}

In this work we have proposed a theoretical framework where neutrino masses are generated at the one-loop level by the exchange of colored particles comprising of vector-like quarks $\Psi$ and two scalar leptoquarks $\eta$ and $\chi$, as shown in Fig.~\ref{fig:neutrinoDirac1loopcolor}. A PQ symmetry under which the exotic fermions $\Psi$ are chirally charged leads to a KSVZ-type axion, solving the strong CP problem.
By construction, the PQ symmetry ensures the Dirac nature of light neutrinos and forbids unwanted terms that could lead to proton decay.  

We examined all possible scenarios resulting in heavy-light quark mixing, specifically the mixing between $\Psi$ and the SM quarks. These enable us to consider the more predictive post-inflationary axion DM picture. In this framework, numerical simulations of axion strings, for $N_{\text{DW}}=1$, restrict the axion scale to the range $5 \times 10^{9}$ GeV to $3 \times 10^{11}$ GeV, in order to account for the full CDM budget. As depicted in Fig.~\ref{fig:gagg} this axion mass region is already partially probed by haloscopes which, together with helioscopes, can scrutinize our various models through their distinctive axion-to-photon coupling predictions. These provide a way to differentiate our various models in future experiments such as IAXO, ADMX, and MADMAX. Moreover, due to the heavy-light quark mixing, some of our models allow for complementary phenomenological features associated to sizable axion-to-quark flavor-violating couplings as summarized in Table~\ref{tab:axionflavorviolating}. This will provide means to constrain them, namely through astrophysical bounds, top decays into Higgs boson, meson mixing, and rare quark flavor processes, as seen in Fig.~\ref{fig:axionfvup}. \\[-.3cm]

In summary, our color-mediated Dirac neutrino mass approach provides a unified framework addressing three open problems of the SM of particle physics: neutrino masses, DM, and the strong CP problem. The idea can also be tested beyond the phenomenology of the axion couplings. For example, by implementing it within extended gauge groups, or by incorporating flavor symmetries, one could get intriguing insights into the flavor puzzle and/or the current flavor anomalies~\cite{Hati:2024ppg}. Last but not least, the post-inflationary axion cosmology scenario should warrant further investigation. Topological defects can play an important role, where for example the decay of axion domain walls and string network could lead to interesting signals at gravitational wave observatories~\cite{Roshan:2024qnv}.

\begin{acknowledgments}
We would like to thank Martin Hirsch for helpful discussions. This research is supported by Fundação para a Ciência e a Tecnologia (FCT, Portugal) through the projects CFTP FCT Unit UIDB/00777/2020 and UIDP/00777/2020, CERN/FIS-PAR/0019/2021 and 2024.02004.CERN, which is partially funded through POCTI (FEDER), COMPETE, QREN and EU. This work is also funded by Spanish grants PID2023-147306NB-I00 (AEI/10.13039/501100011033), Prometeo CIPROM/2021/054 and by Severo Ochoa Excellence grant CEX2023-001292-S (MCIU/AEI/10.13039/501100011033). The work of A.B. and H.B.C. is supported by the PhD FCT Grants No. UI/BD/154391/2023 and 2021.06340.BD, respectively.
\end{acknowledgments}

\appendix

\section{Scalar sector}
\label{sec:scalarsector}

In Tables~\ref{tab:general} and~\ref{tab:postcharges}, we present the particle content and charge assignments of the fields under the SM gauge and PQ symmetries. Besides the Higgs doublet $\Phi$ the scalar sector includes a complex scalar singlet $\sigma$. They are responsible for breaking the electroweak~(EW) and PQ symmetries, through the non-zero VEVs $\langle \phi^0\rangle = v/\sqrt{2} \simeq 174$~GeV and $\langle \sigma\rangle = v_\sigma /\sqrt{2}$, respectively.
Here $\phi^0$ is the neutral component of the Higgs doublet and $\langle \sigma\rangle$ is the PQ breaking scale related to the axion decay constant as shown in Eq.~\eqref{eq:axionfa}. 
There are also two colored scalars $\eta \sim (\mathbf{3}, \mathbf{n}_\Psi \pm \mathbf{1}, y_\Psi + 1/2)$ and $\chi \sim (\mathbf{3}, \mathbf{n}_\Psi, y_\Psi)$, which do not acquire VEVs.
The most general scalar potential allowed by the symmetries of our models is given as
\begin{align}
V & = \sum_S \left[ \mu_{S}^2 S^{\dag}S +\frac{\lambda_S}{4}\left(S^{\dag}S\right)^2\right] + \sum_{S\neq S^\prime}
\frac{\lambda_{SS^\prime}}{2}\left(S^{\dag}S\right)\left({S^\prime}^{\dag}S^\prime\right) + \tilde{\lambda}_{\Phi\eta}\left(\Phi^{\dag}\eta\right)\left(\eta^{\dag}\Phi\right) \nonumber \\
&+ \left(\kappa\, \eta^\dagger \Phi \, \chi + \mathrm{H.c.}\right) \; ,
\label{eq:Vpotential}
\end{align}
where $S,S^\prime = \Phi, \sigma, \eta, \chi$. The cubic term is crucial for neutrino mass generation as discussed in Sec.~\ref{sec:Diracneutrinos}. 
The $\eta$ and $\chi$ (mass eigenstates) with the same electric charge as the colored fermion $\Psi$, namely $Q_{\Psi}$, will mediate Dirac neutrino masses at the one-loop level together with $\Psi$. 
Hence, after spontaneous symmetry breaking~(SSB), the colored scalars with electric charge $Q_{\Psi}$, in the basis $(\eta_{Q_{\Psi}}, \chi_{Q_{\Psi}})$, will mix through,
\begin{align}
    \mathcal{M}_{\eta \chi}^2 = \begin{pmatrix}
        m^2_{\eta} & \frac{\kappa v}{\sqrt{2}} \\
        \frac{\kappa v}{\sqrt{2}} & m^2_{\chi} 
    \end{pmatrix} \; , \;
    m^2_{\eta} &= \mu_{\eta}^2 + \frac{1}{4} \left[(\lambda_{\Phi \eta}+\tilde{\lambda}_{\Phi \eta}) v^2 + \lambda_{\sigma \eta} v_\sigma^2\right] \; , \nonumber \\ 
    m^2_{\chi} &= \mu_{\chi}^2 + \frac{1}{4} (\lambda_{\Phi \chi} v^2 + \lambda_{\sigma \chi} v_\sigma^2) \; .
\end{align}
The corresponding mass eigenstates $\zeta_{1,2}$ are expressed through the $2 \times 2$ rotation matrix $\mathbf{R}$,
\begin{align}
\begin{pmatrix}
        \eta_{Q_{\Psi}} \\ \chi_{Q_{\Psi}}
    \end{pmatrix}= \mathbf{R} \begin{pmatrix}
        \zeta_1 \\ \zeta_2
    \end{pmatrix} = \begin{pmatrix}
        \cos \alpha & -\sin \alpha \\
        \sin \alpha & \cos \alpha
    \end{pmatrix} \begin{pmatrix}
        \zeta_1 \\ \zeta_2
    \end{pmatrix}  \; , \; \tan{2 \alpha} = \frac{\sqrt{2}\,\kappa\, v}{ m^2_\eta - m^2_{\chi} } \; ,
    \label{eq:rotationscalars}
\end{align}
with masses given by,
\begin{align}
    m_{\zeta_{1,2}}^2 = \frac{1}{2} \left[ m_\eta^2 + m_\chi^2 \pm \sqrt{(m_\eta^2 - m_\chi^2)^2 + 2 \kappa^2 v^2} \right] \; .
\end{align}
The remaining mass eigenstates stemming from the weak multiplet $\eta$ ($\chi$) with distinct electric charges are degenerate in mass with value $m_\eta$ ($m_\chi$).

\section{Heavy-light quark mixing}
\label{sec:quarksector}

For all the models in Tables~\ref{tab:general} and~\ref{tab:postcharges} the colored fermion Yukawa Lagrangian contains 
\begin{align}
    -\mathcal{L}_\mathrm{Yuk.}^{\text{c}} \supset -\mathcal{L}_\mathrm{Yuk.}^{\text{c} , 0} = \Y_d \overline{q_L} \Phi d_R + \Y_u \overline{q_L} \Tilde{\Phi} u_R + \mathbf{Y}_\Psi \overline{\Psi_L} \Psi_R \sigma + \mathrm{H.c.} \, ,
    \label{eq:LQuarkYukBase}
\end{align}
the usual SM quark mass terms and that of the VLQs, generated by the PQ breaking scalar. The $3\times 3$ matrix $\Y_{d,u}$ and the $2 \times 2$ matrix $\mathbf{Y}_\Psi$ are in general complex. After SSB, we obtain the following colored fermion mass Lagrangian,
\begin{align}
    -\mathcal{L}_\mathrm{mass}^{\text{c}} \supset -\mathcal{L}_\mathrm{mass}^{\text{c} , 0} & = \overline{d_L} \mathbf{M}_d d_R +  \overline{u_L} \mathbf{M}_u u_R +  \overline{\Psi_L} \mathbf{M}_\Psi \Psi_R + \mathrm{H.c.} \, , \nonumber \\
    \mathbf{M}_d & = \frac{v}{\sqrt{2}} \Y_d \; , \; \mathbf{M}_u = \frac{v}{\sqrt{2}} \Y_u \; ,\; \mathbf{M}_\Psi = \frac{v_\sigma}{\sqrt{2}} \mathbf{Y}_\Psi \; .
    \label{eq:LQuarkYukBase}
\end{align}
In this appendix, we will discuss the specific models presented in Table~\ref{tab:postcharges}, and compute the heavy-light quark mixing, i.e. the mixing between the heavy PQ scale mass VLQ $\Psi$, introduced to solve the strong CP problem and mediate neutrino mass generation, and the ordinary SM quarks with masses proportional to the EW scale.  
In what follows, for simplicity, we discuss the PQ charge assignments $\mathcal{Q}_\text{PQ}$ leading to a few heavy-light mixing terms presented in column 5 of Table~\ref{tab:postcharges}. 
Namely, for all cases we take $\mathcal{Q}_\text{PQ} = 0$, except for $\Psi \sim (\mathbf{3}, \mathbf{2}, 1/6)$ where we take $\mathcal{Q}_\text{PQ} = 1/2$.
The heavy-light mixing is crucial for the computation of the flavor-violating quark-axion couplings, whose phenomenological implications are discussed in Sec.~\ref{sec:axionflavorviolating}.

There are a total of seven models depending on the VLQ representation: 
\begin{itemize}
    \item \underline{$\Psi \sim (\mathbf{3}, \mathbf{1}, -1/3)$:} For this case we have  (see Table~\ref{tab:postcharges}),
\begin{align}
    -\mathcal{L}_\mathrm{Yuk.}^{\text{c}} & =  \mathcal{L}_\mathrm{Yuk.}^{c,0} + \mathbf{M}_{\Psi d} \overline{\Psi_L} d_R + \mathrm{H.c.} \, ,
    \label{eq:Lyuk1down}
\end{align}
where $\mathbf{M}_{\Psi d}$ is a $2 \times 3$ bare mass matrix. Defining, $D_{L,R} = (d,\Psi)^T_{L,R}$, after SSB, we can write the mass Lagrangian in the compact form:
\begin{align}
-\mathcal{L}_\mathrm{mass}^{\text{c}} & = \overline{D_L} \boldsymbol{\mathcal{M}}_d D_R 
    + \overline{u_L} \mathbf{M}_u u_R + \mathrm{H.c.} \; , \; \boldsymbol{\mathcal{M}}_d = \begin{pmatrix}
        \mathbf{M}_d & 0 \\
        \mathbf{M}_{\Psi d} & \mathbf{M}_\Psi
    \end{pmatrix} \; .
    \label{eq:Lcompact1down}
\end{align}
The SM up-quark mass matrix is diagonalised in the standard fashion through the unitary transformations $u_{L,R} \to \mathbf{V}_{L,R}^u\, u_{L,R}$, determined by diagonalising the Hermitian matrices $\mathbf{H}_{u} = \mathbf{M}_{u} \mathbf{M}_{u}^{\dagger}$
and $\mathbf{H}_{u}^\prime = \mathbf{M}_{u}^{\dagger} \mathbf{M}_{u}$. Namely, 
\begin{align}
\mathbf{V}_L^{u \dagger} \mathbf{M}_u \mathbf{V}_R^u = \mathbf{D}_{u} = \text{diag}\,(m_u, m_c, m_t)\,,
\label{eq:diagupSM}
\end{align}
where $m_{u,c,t}$ are the physical light up-type quark masses. The  full quark mass matrix $\boldsymbol{\mathcal{M}}_d$ can be diagonalized through the bi-unitary transformations $(d, \Psi)_{L,R} \to \mathcal{V}_{L,R}^d\, (d, \Psi)_{L,R}$ as 
\begin{align}
\mathcal{V}_L^{d \dagger} \mathcal{M}_d \ \mathcal{V}_R^d = \mathcal{D}_{d} = \text{diag}\,(m_d, m_s, m_b , \tilde{M}_{\Psi_1}, \tilde{M}_{\Psi_{2}})\,,
\label{eq:diagdownfull}
\end{align}
where $m_{d,s,b}$ are the physical light down-type quark masses and $\tilde{M}_{\Psi_{1,2}}$ are the heavy quark masses.
For a given $ \mathcal{M}_d$, $\mathcal{V}_L^d$ and $\mathcal{V}_R^d$ are determined by diagonalising the Hermitian matrices $\mathcal{H}_{d} = \mathcal{M}_{d} \mathcal{M}_{d}^{\dagger}$
and $\mathcal{H}_{d}^\prime = \mathcal{M}_{d}^{\dagger} \mathcal{M}_{d}$. 
 
Since the heavy quark mass scale is set by the PQ breaking scale, in what follows we adopt the seesaw approximation $M_d \ll M_{\Psi d}\,  M_\Psi$.  
Hence, we start by block diagonalising the Hermitian matrix $\mathcal{H}_{d}$, obtaining for the light and heavy sector mass
\begin{equation}
\mathbf{\Lambda}_{d} \simeq \mathbf{M}_d \left(\mathbb{1}_3-\mathbf{M}_{\Psi d}^{\dagger}\mathbf{\Lambda}_{\Psi}^{-1} \mathbf{M}_{\Psi d}\right) \mathbf{M}_d^\dagger \; , \; \mathbf{\Lambda}_{\Psi}^d \simeq \mathbf{M}_\Psi \mathbf{M}_\Psi^\dagger + \mathbf{M}_{\Psi d} \mathbf{M}_{\Psi d}^\dagger \; .
\end{equation}
Furthermore, for $M_{\Psi d} \ll M_\Psi$, $\mathbf{\Lambda}_d$ becomes,
\begin{equation}
\mathbf{\Lambda}_{d} \simeq \mathbf{M}_d \mathbf{M}_d^\dagger - \mathbf{M}_d
\mathbf{M}_{\Psi d}^{\dagger}(\mathbf{M}_\Psi \mathbf{M}_\Psi^\dagger)^{-1} \mathbf{M}_{\Psi d}\mathbf{M}_d^\dagger
\; .
\end{equation}
In this limit, the leading contributions to light down-quark masses come mainly from the SM Yukawa interactions, with the contributions from the heavy-light quark mixing being sub-leading. The matrix $\mathbf{\Lambda}_d$ is diagonalized through the $3 \times 3$ unitary transformation $d_{L} \to \mathbf{V}_{L}^d\, d_{L}$, leading to the well-known Cabibbo-Kobayashi-Maskawa~(CKM) quark mixing matrix $\mathbf{V}_{\text{CKM}} = \mathbf{V}^{u\,\dagger}_L \mathbf{V}^d_L$ appearing in quark charged-current interactions. Furthermore, the heavy sector mass matrix $\mathbf{\Lambda}_\Psi^d$ is diagonalized through the $2 \times 2$ unitary transformation $\Psi_{L} \to \mathbf{V}_{L}^\Psi\, \Psi_{L}$.

The block diagonalisation procedure above was achieved by performing appropriate unitary transformations on the LH quark fields. Similarly, we can diagonalize the Hermitian matrix $\mathcal{H}_d^\prime$ by rotating the right-handed quark fields. We parameterize the unitary transformations $\mathcal{V}_{L,R}^d$ of Eq.~\eqref{eq:diagdownfull}, that relate the weak and mass basis, as follows:
\begin{align}
    \begin{pmatrix}
    \bm{d} \\
    \bm{\Psi}
    \end{pmatrix}_{L,R}
    =  \begin{pmatrix}
    (\mathbb{1}_{3}- \mathbf{\Theta}^d \mathbf{\Theta}^{d \dagger})^{1/2} & \mathbf{\Theta}^d \\
    -\mathbf{\Theta}^{d \dagger} & (\mathbb{1}_{2}-  \mathbf{\Theta}^{d \dagger} \mathbf{\Theta}^d)^{1/2}
    \end{pmatrix}_{L,R}
    \begin{pmatrix}
    \mathbf{V}^d & 0 \\
    0 & \mathbf{V}^\Psi
    \end{pmatrix}_{L,R}
    \begin{pmatrix}
    d\\
   \Psi
    \end{pmatrix}_{L,R} \; .
    \label{eq:qtrans}
\end{align}
The mixing patterns between light and heavy quarks are given by the $3\times 2$ matrices:
\begin{equation}
    \mathbf{\Theta}_R^d \simeq \mathbf{M}_{\Psi d}^\dagger\, \mathbf{M}_\Psi^{\dagger -1}\; , \;
    \mathbf{\Theta}_L^d \simeq \mathbf{M}_d\,\mathbf{M}_{\Psi d}^\dagger\,(\mathbf{\Lambda}_\Psi^{d})^{-1}\simeq \mathbf{M}_d\,\mathbf{\Theta}_R^d\, \mathbf{M}_\Psi^{-1}\,.
    \label{eq:HLmixingrefdown}
\end{equation}
Note that, since $\Theta_L^d \sim \mathcal{O}(v/f_{\text{PQ}}) \Theta_R^d$, with $v/f_{\text{PQ}} \sim \mathcal{O}(10^{-10}) \ll 1$, hence $\mathbf{\Theta}_L^d\ll \mathbf{\Theta}_R^d \ll 1$ and, therefore, we can safely neglect contributions coming from $\mathbf{\Theta}_L^d$. Since $\mathbf{M}_{\Psi d}$ is an arbitrary bare mass term we can choose its scale $M_{\Psi d}$ close to the $M_\Psi \sim \mathcal{O}(f_{\text{PQ}})$ scale, leading to a sizable $\Theta_R^d \sim M_{\Psi d}/M_\Psi$. This will translate into sizable flavor-violating quark-axion couplings as discussed in Sec.~\ref{sec:axionflavorviolating}. 

    \item \underline{$\Psi \sim (\mathbf{3}, \mathbf{1}, 2/3)$:} The quark Yukawa Lagrangian is given by (see Table~\ref{tab:postcharges}),
\begin{align}
    -\mathcal{L}_\mathrm{Yuk.}^{\text{c}} & =  \mathcal{L}_\mathrm{Yuk.}^{c,0} + \mathbf{M}_{\Psi u} \overline{\Psi_L} u_R + \mathrm{H.c.} \, ,
    \label{eq:Lyuk1up}
\end{align}
where $\mathbf{M}_{\Psi u}$ is a $2 \times 3$ bare mass matrix. Defining, $U_{L,R} = (u,\Psi)^T_{L,R}$, after SSB, we can write the mass Lagrangian in the compact form:
\begin{align}
-\mathcal{L}_\mathrm{mass}^{\text{c}} & = \overline{U_L} \boldsymbol{\mathcal{M}}_u U_R 
    + \overline{d_L} \mathbf{M}_d d_R + \mathrm{H.c.} \; , \; \boldsymbol{\mathcal{M}}_u = \begin{pmatrix}
        \mathbf{M}_u & 0 \\
        \mathbf{M}_{\Psi u} & \mathbf{M}_\Psi
    \end{pmatrix} \; .
    \label{eq:Lcompact1up}
\end{align}
We repeat the procedure outlined for the case $\Psi \sim (\mathbf{3}, \mathbf{1}, -1/3)$, making the replacement $d \rightarrow u$ for the up-type sector. Thus, in total analogy, here we can safely neglect contributions coming from $\mathbf{\Theta}_L^u$. However, $\mathbf{\Theta}_R^u$ can lead to sizable axion flavor-violating couplings (see Sec.~\ref{sec:axionflavorviolating}).
    
    \item \underline{$\Psi \sim (\mathbf{3}, \mathbf{2}, 1/6)$:} Defining, $\Psi_{L,R} \equiv (T,B)_{L,R}^T$, the colored fermion Yukawa Lagrangian for this case is (see Table~\ref{tab:postcharges}),
\begin{align}
    -\mathcal{L}_\mathrm{Yuk.}^{\text{c}} & =  \mathcal{L}_\mathrm{Yuk.}^{c,0} + \mathbf{M}_{q \Psi} \overline{q_L} \Psi_R + \mathrm{H.c.} \, ,
    \label{eq:Lyuk1up}
\end{align}
where $\mathbf{M}_{q \Psi}$ is a $3 \times 2$ bare mass matrix. Defining, $D_{L,R} = (d,B)^T_{L,R}$ and $U_{L,R} = (u,T)^T_{L,R}$, after SSB, we obtain,
\begin{align}
-\mathcal{L}_\mathrm{mass}^{\text{c}} & = \overline{D_L} \boldsymbol{\mathcal{M}}_d D_R 
    + \overline{U_L} \boldsymbol{\mathcal{M}}_u U_R 
    + \mathrm{H.c.} \; , \; \nonumber \\
\boldsymbol{\mathcal{M}}_d & = \begin{pmatrix}
        \mathbf{M}_d & \mathbf{M}_{q \Psi}  \\
         0 & \mathbf{M}_\Psi
    \end{pmatrix} \; , \; \boldsymbol{\mathcal{M}}_u = \begin{pmatrix}
        \mathbf{M}_u & \mathbf{M}_{q \Psi} \\
        0 & \mathbf{M}_\Psi
    \end{pmatrix} \; .
    \label{eq:Lcompact2SM}
\end{align}
Repeating the block-diagonalization procedure outlined above, we obtain the mixing patterns between light and heavy quarks for this specific model which are given by the following $3\times 2$ matrices:
\begin{align}
    \mathbf{\Theta}_L^{d,u} \simeq \mathbf{M}_{q \Psi} \, \mathbf{M}_\Psi^{-1}\; , \;
    \mathbf{\Theta}_R^{d,u} & \simeq \mathbf{M}_{d,u}^{\dagger}\,\mathbf{M}_{q \Psi} \,(\mathbf{\Lambda}_\Psi)^{-1} \simeq \mathbf{M}_{d,u}^\dagger\,\mathbf{\Theta}_L^{d,u}\, (\mathbf{M}_\Psi^{\dagger})^{-1} \, , \nonumber \\
\mathbf{\Lambda}_{\Psi} & \simeq \mathbf{M}_\Psi^\dagger \mathbf{M}_\Psi + \mathbf{M}_{q\Psi}^\dagger \mathbf{M}_{q \Psi} \; .
\label{eq:HLmixing2SM}
\end{align}
Note that, for this case, $\mathbf{\Theta}_R^{d,u}$ can be safely neglected while $\mathbf{\Theta}_L^{d,u}$ can be sizable leading to interesting flavor violating axion phenomenology (see Sec.~\ref{sec:axionflavorviolating}). 
    
    \item \underline{$\Psi \sim (\mathbf{3}, \mathbf{2}, -5/6)$:} In this model $\Psi_{L,R} \equiv (B,Y)_{L,R}^T$, with $B$ and $Y$ having electric charge $-1/3$ and $-4/3$, respectively (see Table~\ref{tab:postcharges}). The Yukawa Lagrangian is,
\begin{align}
    -\mathcal{L}_\mathrm{Yuk.}^{\text{c}} & =  \mathcal{L}_\mathrm{Yuk.}^{c,0} + \Y_{\Psi d} \overline{\Psi_L} \tilde{\Phi} d_R + \mathrm{H.c.} \, ,
    \label{eq:Lyuk2down}
\end{align}
where $\mathbf{Y}_{\Psi d}$ is a $2 \times 3$ complex Yukawa matrix. Defining, $D_{L,R} = (d,B)^T_{L,R}$, after SSB, we have,
\begin{align}
-\mathcal{L}_\mathrm{mass}^{\text{c}} & = \overline{D_L} \boldsymbol{\mathcal{M}}_d D_R 
    + \overline{u_L} \mathbf{M}_u u_R + \mathbf{M}_\Psi \overline{Y_L} Y_R  
    + \mathrm{H.c.} \; , \nonumber \\
    \boldsymbol{\mathcal{M}}_d & = \begin{pmatrix}
        \mathbf{M}_d & 0  \\
         \mathbf{M}_{\Psi d} & \mathbf{M}_\Psi
    \end{pmatrix} \; , \; \mathbf{M}_{\Psi d} = \frac{v}{\sqrt{2}} \mathbf{Y}_{\Psi d} \; .
    \label{eq:Lcompact2down}
\end{align}
Block-diagonalizing the above mass matrix leads to the analogous heavy-light mixing patterns as for the $\Psi \sim (\mathbf{3},\mathbf{1},-1/3)$ [see Eq.~\eqref{eq:HLmixingrefdown}]. However, here $\mathbf{M}_{\Psi d}$  is no longer an arbitrary bare mass term but is given by a Yukawa interaction with the Higgs doublet as shown in the above equation.
Consequently, for this case we have $\mathbf{\Theta}_R^d \sim \mathcal{O}(v/f_{\text{PQ}}) \sim \mathcal{O}(10^{-10})$ and $\mathbf{\Theta}_L^d \sim \mathcal{O}(v/f_{\text{PQ}} \mathbf{\Theta}_R^d) \sim \mathcal{O}(10^{-20})$, thus both heavy light-mixing parameters are completely negligible.

\item \underline{$\Psi \sim (\mathbf{3}, \mathbf{2}, 7/6)$:} In this model $\Psi_{L,R} \equiv (X,T)_{L,R}^T$, with $T$ and $X$ having electric charge $2/3$ and $5/3$, respectively (see Table~\ref{tab:postcharges}). We have,
\begin{align}
    -\mathcal{L}_\mathrm{Yuk.}^{\text{c}} & =  \mathcal{L}_\mathrm{Yuk.}^{c,0} + \Y_{\Psi u} \overline{\Psi_L} \Phi u_R + \mathrm{H.c.} \, ,
    \label{eq:Lyuk2up}
\end{align}
where $\mathbf{Y}_{\Psi u}$ is a $2 \times 3$ complex Yukawa matrix. Defining $U_{L,R} = (u,T)^T_{L,R}$ leads to
\begin{align}
-\mathcal{L}_\mathrm{mass}^{\text{c}} & = \overline{d_L} \mathbf{M}_d d_R 
    + \overline{U_L} \boldsymbol{\mathcal{M}}_u U_R + \mathbf{M}_\Psi \overline{X_L} X_R  
    + \mathrm{H.c.} \; , \nonumber \\
    \boldsymbol{\mathcal{M}}_u & = \begin{pmatrix}
        \mathbf{M}_u & 0 \\
        \mathbf{M}_{\Psi u}  & \mathbf{M}_\Psi
    \end{pmatrix} \; , \; \mathbf{M}_{\Psi u} = \frac{v}{\sqrt{2}} \mathbf{Y}_{\Psi u} \; .
    \label{eq:Lcompact2up}
\end{align}
The heavy-light mixing patterns are analogous to the ones for $\Psi \sim (\mathbf{3},\mathbf{1}, 2/3)$ with the replacement $d \rightarrow u$ in Eq.~\eqref{eq:HLmixingrefdown}. However, as shown above, here $\mathbf{M}_{\Psi u}$ is no longer an arbitrary bare mass term but is proportional to the EW scale.
Consequently, for this case we have $\mathbf{\Theta}_R^u \sim \mathcal{O}(v/f_{\text{PQ}}) \sim \mathcal{O}(10^{-10})$ and $\mathbf{\Theta}_L^u \sim \mathcal{O}(v/f_{\text{PQ}} \mathbf{\Theta}_R^u) \sim \mathcal{O}(10^{-20})$, being completely negligible.
 
    \item \underline{$\Psi \sim (\mathbf{3}, \mathbf{3}, -1/3)$:} Defining $\Psi_{L,R} = (T,B,Y)_{L,R}^T$, the quark Yukawa Lagrangian is (see Table~\ref{tab:postcharges})
\begin{align}
    -\mathcal{L}_\mathrm{Yuk.}^{\text{c}} & =  \mathcal{L}_\mathrm{Yuk.}^{c,0} + \Y_{q \Psi} \overline{q_L} \Phi \tau^a \Psi_R^a + \mathrm{H.c.} \, ,
    \label{eq:Lyuk1up}
\end{align}
where $\mathbf{Y}_{q \Psi}$ is a $3 \times 2$ complex Yukawa matrix and $\tau^a$ ($a=1,2,3$) are the Pauli matrices. With $D_{L,R} = (d,B)^T_{L,R}$ and $U_{L,R} = (u,T)^T_{L,R}$, after SSB, 
\begin{align}
-\mathcal{L}_\mathrm{mass}^{\text{c}} & = \overline{D_L} \boldsymbol{\mathcal{M}}_d D_R 
    + \overline{U_L} \boldsymbol{\mathcal{M}}_u U_R 
    + \mathrm{H.c.} \; , \; \nonumber \\
\boldsymbol{\mathcal{M}}_d &= \begin{pmatrix}
        \mathbf{M}_d & -\mathbf{M}_{q \Psi} \\
         0 & \mathbf{M}_\Psi
    \end{pmatrix} \; , \; \boldsymbol{\mathcal{M}}_u = \begin{pmatrix}
        \mathbf{M}_u & \sqrt{2} \mathbf{M}_{q \Psi} \\
        0  & \mathbf{M}_\Psi
    \end{pmatrix} \; , \; \mathbf{M}_{q \Psi} = \frac{v}{\sqrt{2}} \mathbf{Y}_{q \Psi} \; .
    \label{eq:Lcompact3down}
\end{align}
The block-diagonalization procedure leads to:
\begin{align}
\mathbf{\Theta}_L^{d} \simeq - \mathbf{M}_{q \Psi} \, \mathbf{M}_\Psi^{-1}\; , \;
\mathbf{\Theta}_R^{d} & \simeq - \mathbf{M}_{d}^{\dagger}\,\mathbf{M}_{q \Psi} \,(\mathbf{\Lambda}_\Psi^d)^{-1} \simeq - \mathbf{M}_{d}^\dagger\,\mathbf{\Theta}_L^{d}\, (\mathbf{M}_\Psi^{\dagger})^{-1} \, , \nonumber \\
\mathbf{\Lambda}_{\Psi}^d & \simeq \mathbf{M}_\Psi^\dagger \mathbf{M}_\Psi + \mathbf{M}_{q\Psi}^\dagger \mathbf{M}_{q \Psi} \; ; \nonumber \\
\mathbf{\Theta}_L^{u} \simeq \sqrt{2} \mathbf{M}_{q \Psi} \, \mathbf{M}_\Psi^{-1}\; , \;
\mathbf{\Theta}_R^{u} & \simeq \sqrt{2}\mathbf{M}_{u}^{\dagger}\,\mathbf{M}_{q \Psi} \,(\mathbf{\Lambda}_\Psi^u)^{-1} \simeq \sqrt{2} \mathbf{M}_{u}^\dagger\,\mathbf{\Theta}_L^{u}\, (\mathbf{M}_\Psi^{\dagger})^{-1} \, , \nonumber \\
\mathbf{\Lambda}_{\Psi}^u & \simeq \mathbf{M}_\Psi^\dagger \mathbf{M}_\Psi + 2 \mathbf{M}_{q\Psi}^\dagger \mathbf{M}_{q \Psi} \; .
\label{eq:HLmixing3down}
\end{align}
For this case $\mathbf{\Theta}_{R,L}^{d,u}$ are completely negligible.

\item \underline{$\Psi \sim (\mathbf{3}, \mathbf{3}, 2/3)$:} Defining $\Psi_{L,R} = (X,T,B)_{L,R}^T$, we have (see Table~\ref{tab:postcharges}),
\begin{align}
    -\mathcal{L}_\mathrm{Yuk.}^{\text{c}} & =  \mathcal{L}_\mathrm{Yuk.}^{c,0} + \Y_{q \Psi} \overline{q_L} \tilde{\Phi} \tau^a \Psi_R^a + \mathrm{H.c.} \, ,
    \label{eq:Lyuk3up}
\end{align}
where $\mathbf{Y}_{q \Psi}$ is a $3 \times 2$ complex Yukawa matrix and $\tau^a$ ($a=1,2,3$) are the Pauli matrices. Defining, $D_{L,R} = (d,B)^T_{L,R}$ and $U_{L,R} = (u,T)^T_{L,R}$, after SSB, we have,
\begin{align}
-\mathcal{L}_\mathrm{mass}^{\text{c}} & = \overline{D_L} \boldsymbol{\mathcal{M}}_d D_R 
    + \overline{U_L} \boldsymbol{\mathcal{M}}_u U_R 
    + \mathrm{H.c.} \; , \; \nonumber \\
\boldsymbol{\mathcal{M}}_d &= \begin{pmatrix}
        \mathbf{M}_d & - \sqrt{2} \mathbf{M}_{q \Psi}  \\
         0 & \mathbf{M}_\Psi
    \end{pmatrix} \; , \; \boldsymbol{\mathcal{M}}_u = \begin{pmatrix}
        \mathbf{M}_u &  \mathbf{M}_{q \Psi} \\
        0  & \mathbf{M}_\Psi
    \end{pmatrix} \; , \; \mathbf{M}_{q \Psi} = \frac{v}{\sqrt{2}} \mathbf{Y}_{q \Psi} \; .
    \label{eq:Lcompact3down}
\end{align}
The block-diagonalization procedure results in,
\begin{align}
\mathbf{\Theta}_L^{d} \simeq - \sqrt{2} \mathbf{M}_{q \Psi} \, \mathbf{M}_\Psi^{-1}\; , \;
\mathbf{\Theta}_R^{d} & \simeq - \sqrt{2} \mathbf{M}_{d}^{\dagger}\,\mathbf{M}_{q \Psi} \,(\mathbf{\Lambda}_\Psi^d)^{-1} \simeq - \sqrt{2} \mathbf{M}_{d}^\dagger\,\mathbf{\Theta}_L^{d}\, (\mathbf{M}_\Psi^{\dagger})^{-1} \, , \nonumber \\
\mathbf{\Lambda}_{\Psi}^d & \simeq \mathbf{M}_\Psi^\dagger \mathbf{M}_\Psi + 2 \mathbf{M}_{q\Psi}^\dagger \mathbf{M}_{q \Psi} \; ; \; \nonumber \\
\mathbf{\Theta}_L^{u} \simeq \mathbf{M}_{q \Psi} \, \mathbf{M}_\Psi^{-1}\; , \;
\mathbf{\Theta}_R^{u} & \simeq \mathbf{M}_{u}^{\dagger}\,\mathbf{M}_{q \Psi} \,(\mathbf{\Lambda}_\Psi^u)^{-1} \simeq \mathbf{M}_{u}^\dagger\,\mathbf{\Theta}_L^{u}\, (\mathbf{M}_\Psi^{\dagger})^{-1} \, , \nonumber \\
\mathbf{\Lambda}_{\Psi}^u & \simeq \mathbf{M}_\Psi^\dagger \mathbf{M}_\Psi + \mathbf{M}_{q\Psi}^\dagger \mathbf{M}_{q \Psi} \; .
\label{eq:HLmixing3up}
\end{align}
Once again, for this case $\mathbf{\Theta}_{R,L}^{d,u}$ are completely negligible.
    
\end{itemize}
In summary, the models with VLQ $\Psi$ representations $(\mathbf{3}, \mathbf{1}, -1/3)$, $(\mathbf{3}, \mathbf{1}, 2/3)$ and $(\mathbf{3}, \mathbf{2}, 1/6)$, are the only ones where heavy-light quark mixing can be sizable. Specifically, in the RH down quark, RH up quark, and LH up and down quark sectors, respectively. 




\end{document}